\documentclass[letterpaper,twocolumn,10pt]{article}
\usepackage{usenix2019_v3}

\usepackage{tikz}
\usepackage{amsmath}
\usepackage{booktabs}
\usepackage[labelfont=bf,font=small]{caption}
\usepackage{algorithm}
\usepackage{comment}
\usepackage{subcaption}
\usepackage[noend]{algpseudocode}
\usepackage{pseudocode}
\usepackage{blkarray}
\usepackage{xspace}
\algnewcommand{\LineComment}[1]{\State \(\triangleright\) #1}
\usepackage{titling}
\usepackage{enumitem}
\usepackage{microtype}
\usepackage[small,compact]{titlesec}

\usepackage{filecontents}

\newcommand{\system}{Gavel\xspace}

{}

\newcommand{\stan}{{\large$^\dag$}}
\newcommand{\msr}{{\large$^\star$}}

\begin{document}

\date{}

\title{\vspace{-0.3in}\Large \bf Heterogeneity-Aware Cluster Scheduling Policies for Deep Learning Workloads\vspace{-0.2in}}

\author{Deepak Narayanan\stan, Keshav Santhanam\stan, Fiodar Kazhamiaka\stan, Amar Phanishayee\msr, Matei Zaharia\stan \\
\rm{\textit{\msr Microsoft Research\hspace{0.02in} \stan Stanford University}} \\
}

\maketitle

\begin{abstract}

Specialized accelerators such as GPUs, TPUs, FPGAs, and custom ASICs have been
increasingly deployed to train deep learning models.
These accelerators exhibit heterogeneous performance behavior
across model architectures. Existing schedulers for clusters of
accelerators, which are used to arbitrate these expensive training resources across
many users, have shown how to optimize for various \emph{multi-job}, multi-user
objectives, like fairness and makespan. Unfortunately, existing schedulers
largely do not consider performance heterogeneity. In this paper,
we propose Gavel, a heterogeneity-aware scheduler that systematically
generalizes a wide range of existing scheduling policies. Gavel expresses
these policies as optimization problems, making it easy to optimize
for objectives in a heterogeneity-aware way, while also being cognizant of
performance optimizations like space sharing. \system{} then uses a round-based scheduling mechanism
to ensure jobs receive their ideal allocation given the target
scheduling policy.
\system{}'s heterogeneity-aware policies allow a heterogeneous cluster to sustain
higher input load, and improve end objectives such as average
job completion time and makespan by up to 3.5$\times$ compared to
heterogeneity-agnostic policies.

\end{abstract}

\section{Introduction} \label{section:introduction}

As Moore's law comes to an end, specialized accelerators such as GPUs, TPUs, FPGAs,
and other domain-specific architectures have emerged as an alternative to more
general-purpose CPUs. These accelerators have been deployed to great
effect~\cite{jouppi2017datacenter, fowers2018configurable} to train
state-of-the-art deep neural network (DNN) models for many domains, including
language, images and video
~\cite{vaswani2017attention, amodei2016deep, he2017mask, he2016deep,
shafiee2017fast}.

Consequently, users today must choose from a wide variety of accelerators
to train their DNN models. For example, public cloud users can rent several
generations of NVIDIA GPUs and Google TPUs from cloud providers~\cite{aws, gcpgpu, gcptpu}.
Even organizations with private clusters have accumulated different accelerator
types over time~\cite{jeon2019analysis}; anecdotally, our research group has Titan V, Titan X,
and P100 GPUs in its private cluster. Resources in these multi-tenant
settings are typically arbitrated by a scheduler. GPU cluster schedulers such
as Themis~\cite{mahajan2020themis}, Tiresias~\cite{gu2019tiresias},
AlloX~\cite{le2020allox}, and Gandiva~\cite{xiao2018gandiva} thus need to decide how
to allocate diverse resources to many users while implementing complex cluster-wide
\emph{scheduling policies}, optimizing objectives such as fairness or minimum makespan. Unfortunately,
choosing the most effective accelerator types in this context is difficult for three reasons:

\begin{figure}[t!] \centering
  \centering
  \includegraphics[width=0.55\columnwidth]{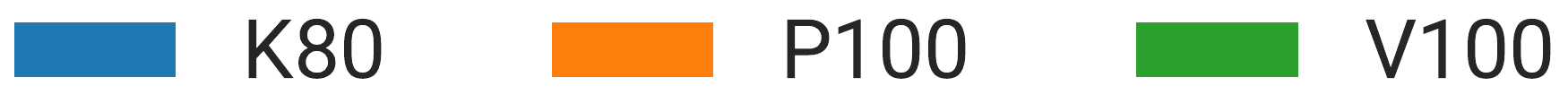}
  \begin{subfigure}[b]{0.9\columnwidth}
    \centering
    \includegraphics[width=1.0\columnwidth]{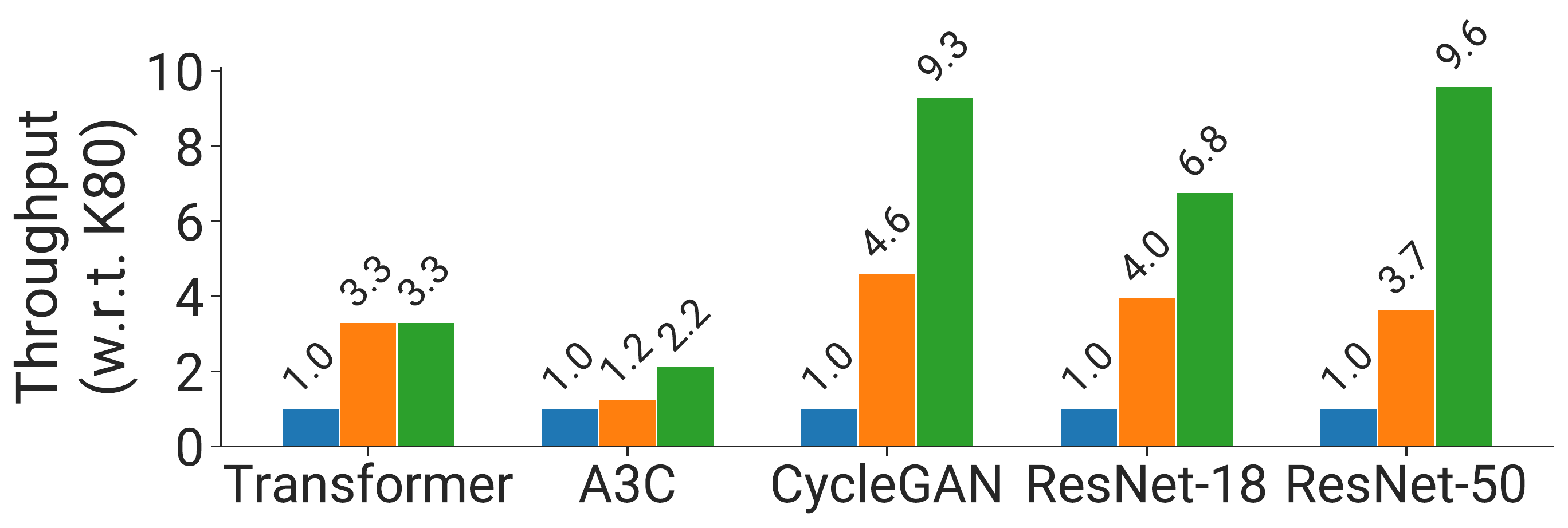}
    \caption{Throughput.}
    \label{fig:heterogeneity}
  \end{subfigure}
  \hspace{0.07in}
  \begin{subfigure}[b]{0.9\columnwidth}
    \centering
    \includegraphics[width=1.0\columnwidth]{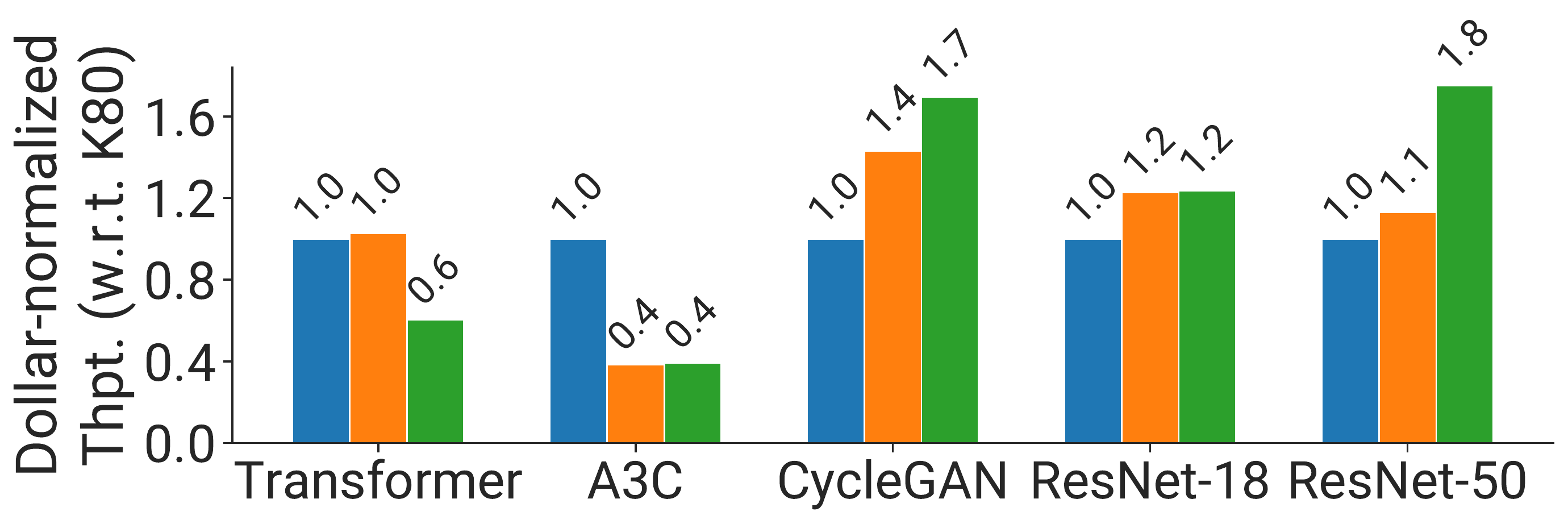}
    \caption{Dollar-normalized.}
    \label{fig:heterogeneity-cost}
  \end{subfigure}
  \caption{Throughputs and dollar-normalized throughputs
  of training for various ML models. Dollar-normalized throughputs are computed by
  dividing the corresponding throughput by the relevant GCP on-demand price,
  The magnitude of speedup across GPU generations
  varies significantly across models.}
  \vspace{-0.15in}
\end{figure}

\paragraph{Performance Heterogeneity.}
Commonly used model architectures show heterogeneous performance behavior across
accelerator types due to various architectural differences.
For example, Figure~\ref{fig:heterogeneity} shows that a ResNet-50 model sees a
nearly 10$\times$ speedup from an NVIDIA V100 GPU compared to a K80 GPU, while an A3C Deep
Reinforcement Learning model only sees a 2$\times$ speedup. However, as shown in
Figure~\ref{fig:heterogeneity-cost}, the V100 is no longer the optimal choice
for all models when we consider the number of samples trained per dollar -- for
many models, the older P100 GPU is competitive or cheaper on a per-dollar basis.
Some scheduling policies can also benefit from splitting a job between \emph{multiple}
resource types: for example, minimizing a job's cost subject to a latency SLO
(e.g., complete a job in 10 hours) might involve using a cheaper accelerator to
begin training and then switching to a faster, more expensive device to meet the SLO.
Thus, for even simple \emph{single-job} settings, the choice of accelerator
type is non-trivial and depends on \emph{both} the job and the policy.
This gets more complicated in \emph{multi-job} settings as granting all jobs their preferred accelerator
simultaneously might not be possible. Existing schedulers like Gandiva, Tiresias,
and Themis do not consider the heterogeneous performance behavior
across accelerators.

\paragraph{Generality Across Policies.}
Cluster operators might want to implement different scheduling policies based
on their business goals, such as optimizing for time to complete
a set of batch jobs (makespan), fairness for ad-hoc jobs, or more sophisticated
\emph{hierarchical} policies that divide resources among high-level entities
(e.g., departments) using one policy, and then individual jobs within
the entity using another~\cite{jeon2019analysis}. In data analytics
clusters, many job schedulers have support for hierarchical allocation
policies~\cite{hadoop-capacity-scheduler, zaharia2010delay,
spark-scheduling-docs,yarn-capacity-scheduler} already.  The two recently
proposed GPU schedulers that do consider heterogeneous resources,
AlloX~\cite{le2020allox} and
$\text{Gandiva}_{\text{fair}}$~\cite{chaudhary2020balancing},
optimize for a single scheduling objective, and tightly couple their scheduling
mechanism to that objective (e.g., max-min fairness). Thus, they cannot
easily support the more sophisticated policies often used in practice.

\paragraph{Colocation and Placement Optimizations.}
To improve cluster utilization, existing GPU schedulers often deploy
optimizations such as space sharing as in Gandiva~\cite{xiao2018gandiva}, where multiple jobs can use the same accelerator
concurrently, and placement sensitivity as in Themis and Tiresias~\cite{mahajan2020themis, gu2019tiresias},
which involves the careful placement of tasks in a distributed job to ensure
good scaling performance. The performance benefits of these optimizations should
be considered explicitly while optimizing for global scheduling objectives, since these
optimizations are more effective when deployed in a heterogeneity-aware way. We
show that this explicit modeling for space sharing can improve objectives by up
to 2.2$\times$ compared to Gandiva's ad-hoc approach.

\vspace{0.2cm}

In this paper, we present \system{}, a new cluster scheduler designed for DNN
training in both on-premise and cloud deployments, that effectively
incorporates heterogeneity in both hardware accelerators and workloads to generalize a
wide range of existing scheduling policies.
For example, \system{} can provide heterogeneity-aware versions
of fair sharing / least attained service~\cite{gu2019tiresias}, FIFO, minimum makespan, minimum cost
subject to SLOs, finish-time fairness~\cite{mahajan2020themis}, shortest job
first, and hierarchical policies~\cite{zaharia2010delay,yarn-capacity-scheduler}.

\system{}'s key observation is that many widely used scheduling policies,
including hierarchical ones, can be expressed as \emph{optimization problems}
whose objective is a function of the jobs' achieved throughputs.
For example, least attained service is equivalent to maximizing the minimum
scaled throughput among the jobs, makespan is equivalent to minimizing the maximum
duration (computed as the ratio of number of iterations to achieved throughput),
and so on. Given the optimization problem for a scheduling policy, Gavel
introduces a general way to transform the problem to make it heterogenity-,
colocation- and placement-aware. In particular, Gavel changes the problem
to search over a heterogeneous \emph{allocation} for each job,
the fraction of time spent in various resource configurations (e.g.,
60\% of time running alone on a V100 GPU and 40\% of time space-sharing an A100
GPU with another job), and changes the throughput terms in the objective
function to \emph{effective throughput}, i.e. the average throughput of the job
over the mix of resources in its allocation. Additional constraints need to be
added to ensure that the returned allocation is valid.
We show that \system{}'s transformed optimization problems are efficient to
execute even for clusters with hundreds of GPUs and jobs, and can support a
wide range of policies. Many of these problems can be solved using a sequence
of one or more linear programs.

\system{}'s heterogeneity-aware allocations for each job need to be mapped
to actual scheduling decisions (placement of jobs on specific resources in the
cluster for a specified duration of time). To achieve this, \system{} uses a
preemptive \emph{round-based scheduling mechanism} to ensure that jobs receive
resources in fractions similar to the computed target allocation. \system{}'s
scheduling mechanism needs to be able to schedule both distributed training
jobs, which request multiple accelerators at once, as well as combinations of
jobs running concurrently on a given accelerator due to space sharing.

\system{} makes these scheduling decisions transparently: it specifies an API
between the scheduler and applications that allow jobs written in existing deep
learning frameworks like PyTorch~\cite{pytorch} and TensorFlow~\cite{tensorflow} to be moved between resources
with minimal code changes, and uses a mechanism similar to Quasar~\cite{delimitrou2014quasar} to estimate
performance measurements of colocated jobs, which are needed as
inputs to \system{}'s policies, when not available \textit{a priori}.

By explicitly considering performance heterogeneity, \system{} improves
various policy objectives (e.g., average job completion time or makespan):
on a smaller physical cluster, it improves objectives
by up to 1.4$\times$, and on a larger simulated cluster, it increases the maximum
cluster load, while improving objectives such as average
job completion time by 3.5$\times$, makespan by 2.5$\times$, and cost by 1.4$\times$.

To summarize, our main contributions are:
1) A systematic method to convert existing cluster scheduling policies into
      equivalent policies that consider heterogeneity and colocation; these
      equivalent optimization problems are practical to
      execute for current DNN clusters.
2) A round-based scheduling mechanism to ensure that the cluster realizes
      the allocations returned by these policies.
3)  Generalizations of many existing policies in our 
      framework that improve overall performance and other policy objectives
      by up to 3.5$\times$.

\begin{figure*}[t]
  \centering
  \includegraphics[width=0.88\textwidth]{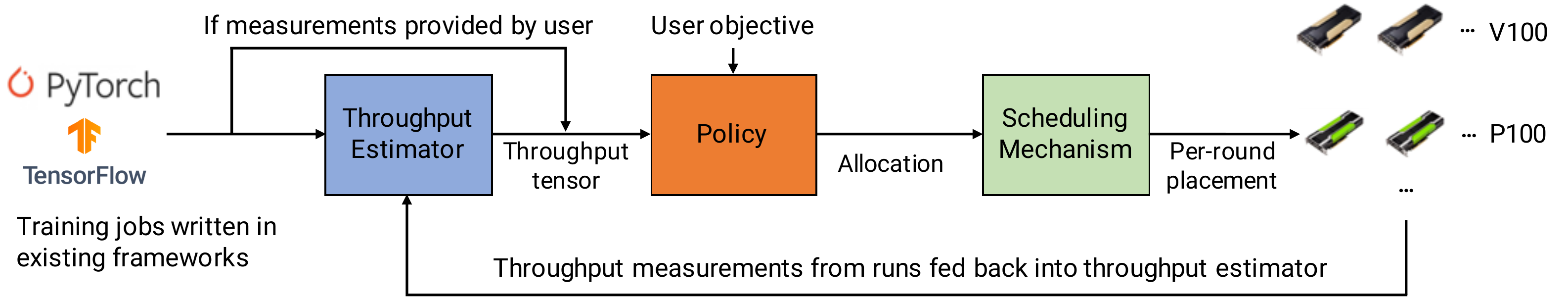}
  \caption{\system{} overview. Jobs are written in frameworks like PyTorch
           or TensorFlow. \system{}'s throughput estimator obtains performance
           measurements for each runnable job on each available accelerator type
           if necessary; its policy then computes an allocation that optimizes a
           user-specified objective such as fairness. \system{}'s scheduling
           mechanism accepts this computed allocation as an input, and makes
           per-round placement decisions in proportions that faithfully mimic
           the computed allocation.}
  \label{fig:system_architecture}
  \vspace{-5mm}
\end{figure*}

\section{Background}

In this section, we provide a brief overview of DNN training
(\S\ref{section:background_dnn}), and discuss performance optimizations used in
existing schedulers that \system{} can help deploy more effectively
(\S\ref{section:background_optimizations}).

\subsection{Deep Neural Network (DNN) Training} \label{section:background_dnn}
DNN training proceeds in iterations. In each iteration, the DNN processes
a collection of inputs (called a minibatch) and subsequently updates the model
parameters using gradients derived from the input minibatch. Each minibatch
is typically of similar size, which means model training throughput can be measured by
averaging over 100s of iterations. Gavel leverages this fact in its throughput
estimator. Jobs are typically fairly long-running (on the order of hours to
days), and can be distributed over many workers~\cite{jeon2019github,
xiao2018gandiva}.

Modern DNN schedulers leverage the fact that DNN training is iterative to suspend
and resume training at iteration boundaries~\cite{gu2019tiresias,xiao2018gandiva};
this ensures that jobs can be time multiplexed over the existing physical
resources. The latest model parameters need to be checkpointed to stable storage
when a job is suspended to ensure training progress is not lost.

\subsection{Performance Optimizations} \label{section:background_optimizations}

Prior work has shown that GPUs can be severely under-utilized in multi-tenant
clusters~\cite{jeon2019analysis}; for example, average GPU utilization (measured
as the percentage of GPU Streaming Multiprocessors active over time) was as low
as $52\%$ on a Microsoft cluster. Prior work has also shown the placement of tasks
for a distributed training job can have significant impact on performance.
\system{} can more systematically employ these optimizations, as we show in
\S\ref{sec:policies_overview}.

\paragraph{Space Sharing.} Smaller models often do not leverage the full
computational capacity of modern GPUs. In such cases, concurrently executing
multiple models on the same GPU using NVIDIA's Multi Process Service (MPS) or
CUDA streams can help improve utilization~\cite{mps, narayanan2018accelerating}.

\paragraph{Placement Sensitivity.} DNN models show heterogeneity in their distributed scaling behavior depending on the
size of the data that need to be exchanged between workers during training: some models
have compact weight representations and can
scale well even when workers are not on the same server, while other
models scale poorly when workers are spread over many servers. Existing
schedulers like Tiresias use heuristics for placement sensitivity to ensure that
model training is run in a consolidated setting when necessary.

\section{Overview of \system{}} \label{section:system_design}

Given a collection of jobs, \system{} arbitrates cluster resources (in the form
of accelerators of different types) among the resident jobs, while optimizing
for the desired cluster objective. This is accomplished in a two-step process:
first, a \emph{heterogeneity-aware policy} computes the fraction of time
different jobs (and combinations) should run on different accelerator types to
optimize the desired objective. These policies require as input the performance
behavior (in terms of throughputs) for each job on each accelerator type, which
can either be provided by the user, or can be measured on the fly by \system{}'s
throughput estimator. Then, given the policy's output allocation, \system{}'s
\emph{scheduling mechanism} grants jobs time on the different resources,
and moves jobs between workers as necessary to ensure that the true fraction
of time each job spends on different resources closely resembles the optimal
allocation returned by the policy. \system{} can recompute its policy either
when a \emph{reset event} occurs (job arrives or completes, or a worker in the
cluster fails), or at periodic intervals of time. \system{}'s workflow is
shown in Figure~\ref{fig:system_architecture}.

\subsection{Heterogeneity-Aware Policies}
\label{sec:policies_overview}

\system{} expresses scheduling policies as optimization problems for various
objectives of interest, such as fairness or makespan, and allocations as matrices that specify the fraction of
wall-clock time a job should spend on each accelerator type.
A matrix $X$ can represent allocations on a
single accelerator type (homogeneous setting), allocations on multiple
accelerator types (heterogeneous setting), as well as allocations with other
optimizations deployed.
For three jobs and three accelerator types, an example
allocation could be:
$$X^\text{example} =
\begin{blockarray}{cccc}
V100 & P100 & K80 \\
\begin{block}{(ccc)c}
  0.6 & 0.4 & 0.0 & \text{job 0} \\
  0.2 & 0.6 & 0.2 & \text{job 1} \\
  0.2 & 0.0 & 0.8 & \text{job 2} \\
\end{block}
\end{blockarray}$$

\begin{figure}[h!]
    \vspace{-0.3in}
    \centering
    \includegraphics[width=0.75\columnwidth]{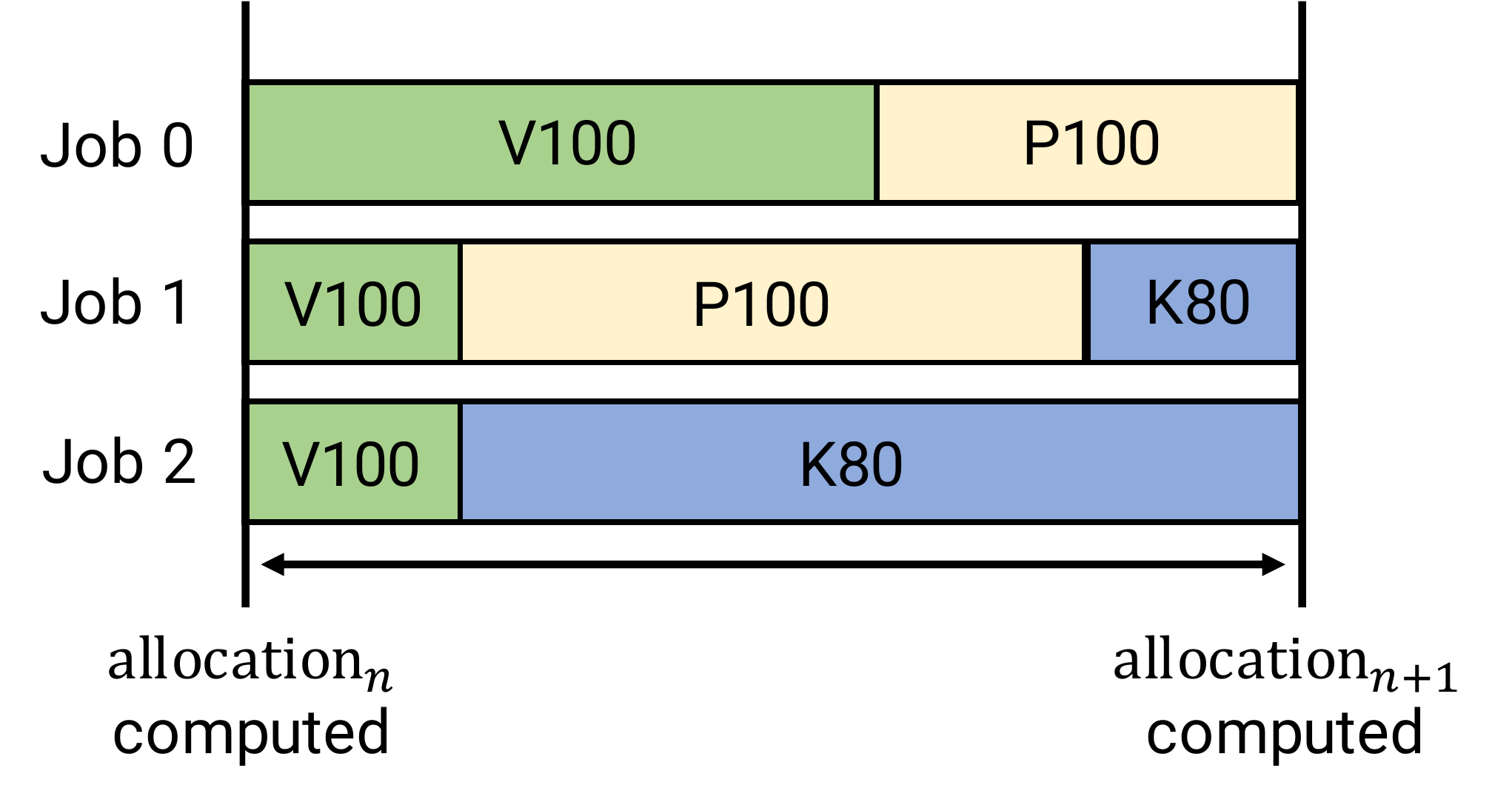}
    \caption{
        \label{fig:allocation}
        The \emph{cumulative} time each job spends on
        accelerator types between allocation recomputations for allocation $X^\text{example}$.
    }
\end{figure}
According to this allocation, job 0 should spend 60\% of its time on a V100 GPU,
and job 1 should spend 40\%. This is shown visually in Figure~\ref{fig:allocation}.

\system{} finds an optimal value for the matrix $X$ given a policy expressed as
an optimization problem. To construct the optimization problem for a given policy,
\system{} requires a \emph{throughput matrix} $T$ with each job's throughput (in training
iterations per second) on different accelerators. $T_{mj}$ can be set to
$-\infty$ if job $m$ does not run on accelerator type $j$ (for example, due to
memory constraints).

Given $T$ and $X$, we define the \emph{effective throughput} of a model $m$
as the time-weighted average throughput across accelerators and jobs. We denote
this quantity $\text{\text{throughput}}_T(m, X)$ or simply
$\text{\text{throughput}}(m, X)$ (dropping the $T$) for brevity. For allocations
$X$ without space sharing,
$$\text{throughput}(m, X) = \sum_{\substack{j \in \\ \text{accelerator types }}} T_{mj} \cdot X_{mj}$$
Different cluster scheduling policies can be expressed as optimization
problems for $X$ while maximizing or minimizing an appropriate objective
function. Constraints need to be specified to ensure that $X$ is a valid
allocation. Then, a hypothetical policy that maximizes total
effective throughput would look like,
$$\text{Maximize}_X \sum_{m \in \text{jobs}} \text{throughput}(m, X)$$
\noindent We need to add additional constraints to ensure that the cluster is not
overprovisioned:
\begin{align}
& 0 \leq X_{mj} \leq 1 & \forall (m,j)  \\
& \sum_j X_{mj} \leq 1 & \forall m  \\
& \sum_m X_{mj} \cdot \text{scale\_factor}_m \leq \text{num\_workers}_j & \forall j \label{con-oversubscribe}
\end{align}
These constraints ensure that each job-worker allocation is non-negative and
is between $0$ and $1$, that the total allocation for a job does not exceed 1,
and that the allocation does not oversubscribe the workers.

\paragraph{Space Sharing.}
\system{}'s allocation matrices can also incorporate space sharing options.
While previous work has used greedy algorithms for space sharing,
we found that different pairs of DNN applications in practice have vastly
different performance when colocated together, based on the resources they consume (Figure~\ref{fig:heatmap} in Appendix).
When using space sharing, $X$ needs to contain rows for each viable
combination of jobs, and $T$ needs to have throughputs of the job combinations, like:
$$T =
\begin{blockarray}{cccc}
V100 & P100 & K80 \\
\begin{block}{(ccc)c}
  4.0 & 2.0 & 1.0 & \text{job 0} \\
  3.0 & 2.0 & 1.0 & \text{job 1} \\
  (2.0, 1.5) & 0.0 & 0.0 & \text{jobs (0, 1)} \\
\end{block}
\end{blockarray}$$
We limit entries of $T$ to combinations of at most 2 jobs; we found empirically that
larger combinations rarely increase net throughput. Additionally, although the size of $T$ grows quadratically
with the number of jobs even with job combinations of size 2, we found that in practice
we only need to consider combinations that actually perform well. We evaluate the scaling behavior
of these SS-aware policies in \S\ref{sec:evaluation_scalability}.

Objectives in terms of $\text{throughput}(m, X)$ remain the same;
however, $\text{throughput}(m, X)$ now needs to be computed to include the throughputs of
co-located jobs, and additional constraints need to be specified to ensure that $X$
is a valid allocation in this new regime:
\begin{eqnarray}
& 0 \leq X_{kj} \leq 1 & \forall k, j \nonumber \\
& \sum_{k \in C_m} \sum_j X_{kj} \leq 1 & \forall m \nonumber \\
& \sum_{k} X_{kj} \cdot \text{scale\_factor}_m \leq \text{num\_workers}_j & \forall j \nonumber
\end{eqnarray}
$C_m$ is the set of all job combinations that contain job $m$.

\paragraph{Placement Sensitivity.}
Similarly, \system{}'s allocation matrices can also be extended to incorporate placement
sensitivity. The observed throughput for distributed jobs depends on the location of tasks,
as well as the model and accelerator type (slower
workers are less likely to be communication-bound, which means consolidation of tasks is
less effective). We can make our policies \emph{placement-sensitive} by considering
the performance of distributed jobs in: 1) a consolidated setting,
where as many accelerators are on the same server as possible (for example, 8 GPUs per
server if using 8-GPU servers), and 2) an unconsolidated setting, where accelerators
are on independent servers. These are extreme points in the placement
space, and are upper and lower bounds on performance. We can model this
in our policies by having two different worker types (consolidated and
unconsolidated) with corresponding throughput values in $T$ and allocation values in $X$.

\begin{figure}
    \centering
    \includegraphics[width=1.0\columnwidth]{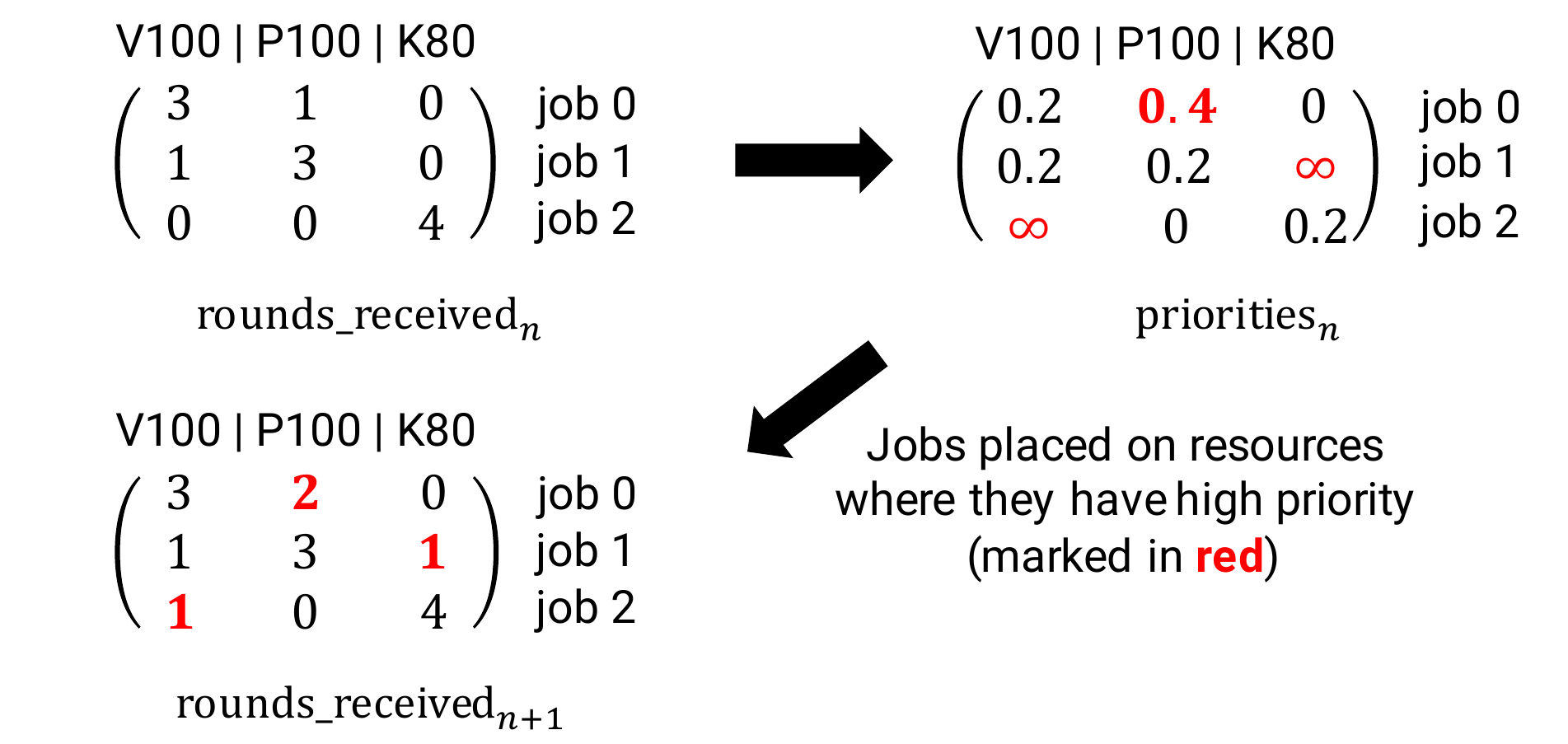}
    \caption{
        \label{fig:priority_mechanism}
        Priorities are used to move the received allocation towards the intended
        allocation (in this case, $X^\text{example}$). $\text{priorities}_n$ is computed as
        $f(X, \text{rounds\_received}_n)$ where $f(.)$ is the element-wise division
        operator.
    }
    \vspace{-5mm}
\end{figure}

\subsection{Round-based Scheduling Mechanism}
After computing the optimal allocation, \system{}'s next step is to assign jobs
(or job combinations, in the case of space sharing) to accelerator types while
matching the optimal allocation as closely as possible.
That is, to realize the allocation $X^\text{example}$ above,
the scheduling mechanism needs to make sure that in the time period where
jobs 0, 1, and 2 are the only three runnable jobs in the cluster, jobs should
receive resources according to their computed fractions (e.g., job 0 receives 60\% of
V100 time).

To do this, the scheduler computes a priority score for every job and accelerator
type combination that is high when a job has received a smaller time fraction
than the optimal allocation. Scheduling is performed in rounds; in each round,
the scheduler runs jobs in decreasing priority order, while ensuring that a given job is
not scheduled on multiple workers (or accelerators) in a given round. This is shown in
Figure~\ref{fig:priority_mechanism}. Priorities are updated as rounds complete.
We have found empirically that round durations of 20 minutes or less allow
\system to effectively approximate the ideal allocation. We present an
analysis of how the round duration affects scheduling quality
and overhead in \S\ref{sec:evaluation_mechanism}.

\subsection{Throughput Estimator}
To estimate the throughputs of concurrent jobs (e.g., in the case of space
sharing), Gavel employs a throughput estimator, similar to those found in prior
work such as Quasar~\cite{delimitrou2014quasar}.
\system{}'s throughput estimator maps
a new job to a set of pre-profiled reference jobs. The throughputs of the
closest reference job can then be used as the initial performance estimate
for the new job's combinations. For individual jobs, the throughput estimator
is not needed, since these can be estimated on the fly as jobs run on different
resource types over many rounds.

\subsection{Limitations and Non-Goals}
While \system{} exposes a flexible API that supports a variety of policies and
objectives, we do not propose new scheduling policies or performance
optimizations in this work. Instead,
\system{}'s main goal is to determine how best to share resources amongst many
different users and jobs in a heterogeneity-aware way, while
supporting \emph{many} existing cluster-wide objectives. \system{} accomplishes
these goals with a policy framework that easily allows policies to
be made heterogeneity-, colocation-, and placement-aware (\S\ref{sec:policies}), a scheduling
mechanism that can be reused across policies (\S\ref{sec:scheduler}), and a narrow scheduler API
that allows users to deploy their applications on the shared
cluster with minimal code changes (\S\ref{sec:implementation}).

\section{Scheduling Policies} \label{sec:policies}

In this section, we show how various scheduling policies such as max-min fairness (Least Attained Service or LAS)
and multi-level fairness can be expressed as optimization problems in terms of effective throughput.
We also describe some properties of the resulting heterogeneity-aware allocations
at the end of this section.

\subsection{Max-Min Fairness as an Optimization Problem}
\label{sec:max_min_fairness}

The classical Least Attained Service (LAS) policy, used by
Tiresias~\cite{gu2019tiresias}, implements max-min fairness across active users
in the cluster, by round-robining resources across jobs according to
the total number of accelerator hours consumed. This can be modified into a
weighted max-min fairness policy with per-user weights $w_m$.

Thus, on a homogeneous cluster, if a job $m$ with weight $w_m$ receives a
fraction $X_m$ (which is a scalar since there is only one resource type), LAS
can be expressed as the following optimization problem:
$$\text{Maximize}_X \min_m \dfrac{1}{w_m}X_m$$
We need to add an additional constraint to ensure that the cluster is not
overprovisioned ($\sum_m X_m \leq 1$).

However, this vanilla LAS policy is not fair in a heterogeneous
setting; jobs might see unequal reductions in throughput due to
variations in performance across accelerator types. For example,
giving one job a K80 and another job a V100 would equalize their number of resources, but
could result in very low performance for the job with the K80.

To compute a more fair allocation, we can compute max-min fairness over the
weighted normalized effective throughputs, as defined in \S\ref{sec:policies_overview}. Let $X_m^\text{equal}$ be the allocation
given to job $m$ assuming it receives equal time share on each worker in the
cluster. For example, if the cluster had 1 V100 and 1 K80,
$X_m^\text{equal} = [0.5, 0.5]$. $X_m^\text{equal}$
scales the effective throughputs to make them comparable across jobs.
$$\text{Maximize}_X \min_m \dfrac{1}{w_m} \dfrac{\text{throughput}(m, X)}{\text{throughput}(m, X_m^{\text{equal}})}$$
As specified in \S\ref{sec:policies_overview}, additional constraints need to
be specified to ensure that allocations are valid.

As an example, consider 3 jobs which benefit differently when moved from
a K80 GPU to a V100 GPU:
$$T =
\begin{blockarray}{ccc}
V100 & K80 \\
\begin{block}{(cc)c}
  4.0 & 1.0 & \text{job 0} \\
  3.0 & 1.0 & \text{job 1} \\
  2.0 & 1.0 & \text{job 2} \\
\end{block}
\end{blockarray}$$
Solving the above optimization problem with $w_m=1$, and a cluster with 1 V100 and 1 K80 yields the
following allocation:
$$X^\text{het.} = \begin{blockarray}{ccc}
V100 & K80 \\
\begin{block}{(cc)c}
  0.45 & 0.0 & \text{job 0} \\
  0.45 & 0.09 & \text{job 1} \\
  0.09 & 0.91 & \text{job 2} \\
\end{block}
\end{blockarray}$$
Jobs receive about 10\% higher throughput compared to
an allocation where every user is given $1/n$ of the time on each accelerator (in this case, $n=3$), also
called an \emph{isolated allocation}~\cite{ghodsi2011dominant}.

Fairness policy objective functions need to be modified to
take into account muti-resource jobs with $\text{scale\_factor}_m > 1$, since these multi-resource jobs
occupy a larger share of the cluster per unit time. An easy way to do this
is to multiply the $\text{scale\_factor}_m$ to the max-min objectives from before.
Concretely, the LAS objective from before now becomes,
$$\text{Maximize}_X \min_m \dfrac{1}{w_m} \dfrac{\text{throughput}(m, X)}{\text{throughput}(m, X_m^{\text{equal}})} \cdot \text{scale\_factor}_m$$

\subsection{Other Policies as Optimization Problems}

\begin{table}[t!]
\footnotesize
\centering
\begin{tabular}{ll}
\toprule
\textbf{Policy} & \textbf{Description} \\
\toprule
Makespan & Minimize time taken by batch of jobs. \\
LAS~\cite{gu2019tiresias} & Max-min fairness by total compute time.\\
LAS w/ weights & Max-min fairness with weights. \\
Finish Time Fairness~\cite{mahajan2020themis} & Maximize minimum job speedup.\\
FIFO & First in, first out. \\
Shortest Job First & Minimize time taken by shortest job. \\
Minimize cost & Minimize total cost in public cloud. \\
Minimize cost w/ SLOs & Minimize total cost subject to SLOs. \\
Hierarchical~\cite{zaharia2010delay} &  Multi-level policy: FIFO, fairness, etc. \\ \bottomrule
\end{tabular}
\caption{Policies that can be expressed in \system{}.}
\label{table:policies}
\end{table}

We can express many other common cluster scheduling policies, some proposed by
recent papers, using $\text{\text{throughput}}(m, X)$; we list these policies in
Table~\ref{table:policies}. Most of these policies can be expressed using a
single linear program, with a few exceptions: the makespan policy is formulated as a sequence of
linear programs, and the cost policies are formulated as a linear-fractional
program~\cite{linearfractionalprogram}, which can be reduced to a sequence of linear programs as well.
These optimization problems yield corresponding heterogeneity-aware allocations.
The optimal allocation can be computed by solving one or more
linear programs (LP), which off-the-shelf solvers can solve very quickly even
at the scale of a large cluster.

\paragraph{Minimize Makespan.}

The Minimum Makespan policy tries to complete all active jobs as soon as
possible. Gandiva uses a version of this policy to finish higher-level tasks
such as hyperparameter tuning and AutoML, which involve training a large number
of variants of a model. If $\text{num\_steps}_m$ is the number of iterations
remaining to train model $m$, then the makespan is the maximum of the durations
of all active jobs, where the duration of job $m$ is the ratio of the number of
iterations to $\text{throughput}(m, X)$ (which is expressed in iterations /
second). Overall, this can be framed as,

$$\text{Minimize}_X \max_m \dfrac{\text{num\_steps}_m}{\text{throughput}(m, X)}$$

\paragraph{Minimize Finish-Time Fairness (Themis).}
Themis~\cite{mahajan2020themis} proposes a new metric called finish-time
fairness (represented as $\rho$), which is the ratio of the time taken to finish
a job using a given allocation and the time taken to finish the job using $1/n$
of the cluster ($X^{\text{isolated}}$), assuming $n$ users using the cluster.
This can be expressed in terms of $\text{throughput}(m, X)$ as follows
($\text{num\_steps}_m$ is the number of iterations remaining to train model $m$,
$t_m$ is the time elapsed since the start of training for model $m$, and
$t^\text{isolated}_m$ is the hypothetical time elapsed since the start of training if model $m$
had a dedicated fraction of the cluster to itself),
$$\rho_T(m, X) = \dfrac{t_m + \frac{\text{num\_steps}_m}{\text{throughput}(m, X)}}{t^\text{isolated}_m + \frac{\text{num\_steps}_m}{\text{throughput}(m, X^{\text{isolated}})}}$$
The final optimization problem is then,
$$\text{Minimize}_X \max_m \rho_T(m, X)$$

\paragraph{FIFO.} The First-In-First-Out (FIFO) policy schedules jobs in
the order they arrive. In a heterogeneous regime, jobs should be placed on the
fastest available accelerator type. Mathematically, we can write this as maximizing
the throughput of job $m$ relative to its throughput on the fastest type
($\text{throughput}(m, X^{\text{fastest}})$). Assuming that jobs are enumerated
in order of their arrival time ($m$ arrived before $m+1$), a FIFO allocation
can be computed with the following objective:

$$\text{Maximize}_X \sum_m \frac{\text{throughput}(m,X)}{\text{throughput}(m,X^{\text{fastest}})}(M-m)$$
where $M$ is the total number of jobs.

\paragraph{Shortest Job First.}

The Shortest Job First policy finds the allocation that minimizes the
duration of the shortest job,

$$\text{Minimize}_X \min_m \dfrac{\text{num\_steps}_m}{\text{throughput}(m, X)}$$

\paragraph{Minimizing Total Cost and Cost subject to SLOs.} \label{section:min_total_cost}

We can express policies for deployments that use elastic public cloud resources.
Since cloud VMs are charged on a per-time basis, we can express policies that
explicitly optimize for total cost, speed, or both.

Consider a simple policy that maximizes total throughput,
$$\text{Minimize}_X \sum_m \text{throughput}(m, X)$$

The above policy can be extended to incorporate cost by optimizing the
following cost-adjusted objective,
$$\text{Maximize}_X \frac{\sum_m \text{throughput}(m, X)}{\sum_m (\sum_j \text{cost}_j \cdot X_{mj})}$$
where $\text{cost}_j$ is the cost of accelerator type $j$. The numerator in the
above objective is the time-averaged effective throughput, and the denominator is
the time-averaged cost. When using space sharing, care must be taken to not
double count the cost of instances running job combinations (all jobs in a job
combination derive value in terms of some throughput from the instance).

Jobs can have time SLOs as well, e.g., certain high-priority jobs might need to
complete every 12 hours. We can add additional constraints: given $\text{SLO}_m$
for each model $m$ (models without SLOs can have $\text{SLO}_m = \infty$),
$$\text{throughput}(m, X) \geq \text{num\_steps}_m / \text{SLO}_m$$

\subsection{Hierarchical Scheduling Policies}

\begin{figure}
    \centering
    \includegraphics[width=0.75\columnwidth]{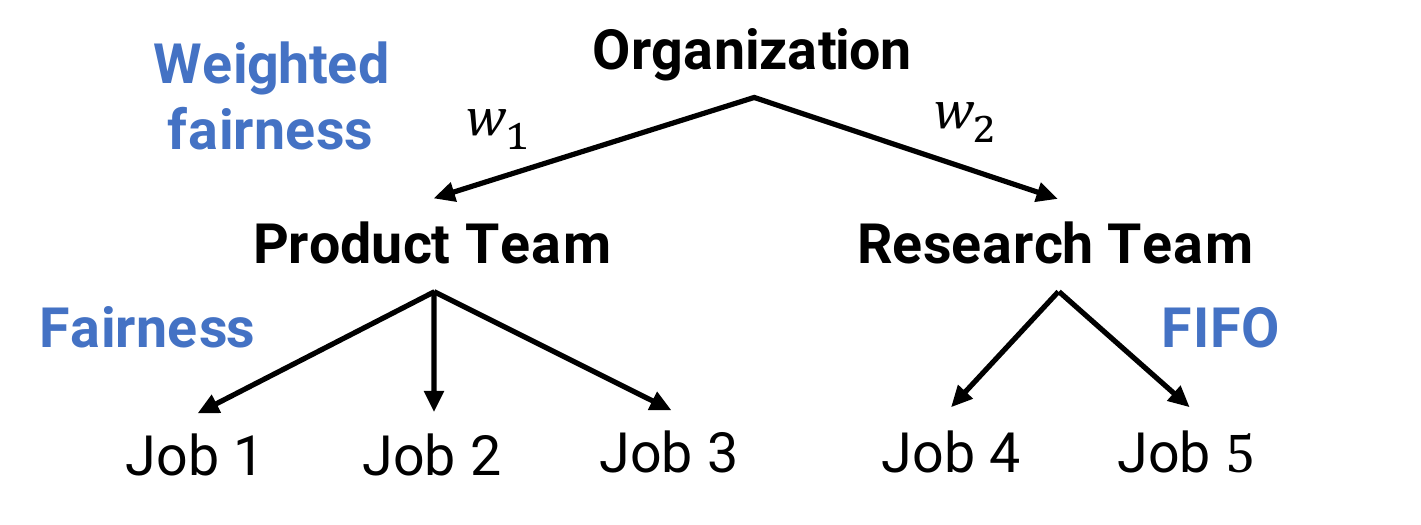}
    \caption{
        \label{fig:hierarchical_fairness}
        Example of a hierarchical policy: weighted fairness across two entities:
        a product and research team, fairness across jobs within the product
        team, and FIFO within the research team.
    }
    \vspace{-5mm}
\end{figure}

Modern cluster schedulers do not only deploy ``single-level'' policies.
Hierarchical policies are common~\cite{hadoop-capacity-scheduler,
zaharia2010delay,yarn-capacity-scheduler}: a large organization might share a single
physical cluster among many sub-organizations (or entities) using a fairness
policy. In turn, each entity can share resources among individual jobs
according to a distinct per-entity policy, such as per-user fairness or FIFO.
We give an example in Figure~\ref{fig:hierarchical_fairness},
where a research and product team share the same physical cluster. The
research team runs ad-hoc experiments that can be executed in FIFO order, but
the product team needs to ensure that all its jobs receive a fair share
of the cluster.

\system{} can currently support fairness in the upper levels and fairness or FIFO
in the lower levels, which matches the hierarchical policies supported by the
Hadoop scheduler~\cite{hadoop-capacity-scheduler}.
Determining how to extend this to other hierarchical policy sets (for example,
with finish time fairness) is future work.

\system{} solves hierarchical objectives using a procedure called water
filling \cite{bertsekas1987data}, which is used in other max-min fairness
problems such as link allocation in networks~\cite{radunovic2007unified}.
At a high level, the water-filling algorithm increases the allocation given
to all parties at an equal rate to respect max-min fairness, until a party saturates.
The saturated party is then taken out, and the procedure repeated iteratively
until all commodities are saturated.
We adapt this procedure to our setting, solving a series of optimization problems iteratively:
an LP that computes a fair allocation across entities while respecting each
entity's internal policy, and an MILP (specified in
Appendix~\ref{app:sec-policy_extensions_multistep}) that identifies \emph{bottlenecked
jobs}, i.e., jobs whose effective throughputs cannot be improved without
lowering the effective throughput of other jobs.

We denote each entity $s$ associated with a weight $w_s$; the jobs
belonging to this entity receive a total share of the cluster proportional
to this weight. We denote $w^\text{job}_m$ to be the weight of job $m$, set such
that $\sum_{m \in s} w^\text{job}_m = w_s$. Jobs are assigned priorities in accordance to the
relevant entity's policy; for example, a fairness policy at the entity level
would assign each job a weight proportional to its individual weight
within the entity, while for FIFO, the first job in the queue would initially
receive the entire weight of the entity.

In each iteration, we solve the following modified LP, assuming $\text{scale\_factor}_m = 1$
for all $m$ for simplicity:
$$\text{Maximize}_X \min_{\{m : w^\text{job}_m > 0\}}\frac{1}{w^\text{job}_{m}} \bigg(\frac{\text{throughput}(m,X)}{\text{throughput}(m,X_m^{\text{equal}})} - t_m \bigg)$$
$t_m$ is the normalized effective throughput
of job $m$ in the previous iteration ($t_m := 0$ in the first iteration).
The above objective can be appropriately modified for larger $\text{scale\_factor}_m$.
Bottlenecked jobs are given priority 0 and no longer
considered in future iterations. Priorities are redistributed among
non-bottlenecked jobs according to the entity's policy at the end of every
iteration. For instance, in the example shown in Figure~\ref{fig:hierarchical_fairness}, 
if job 4 is bottlenecked then its weight is reassigned to job 5 in accordance to the FIFO policy, 
while if job 2 is bottlenecked, its weight is distributed equally between jobs 1 and 3 in accordance with
the entity's fairness policy.
The LP then solves the max-min problem on the resources remaining while ensuring
each job's throughput does not drop compared to the previous iteration's
allocation $X^\text{prev}$, expressed as the constraint
$\text{throughput}(m,X) \geq \text{throughput}(m,X^{\text{prev}})$ for all m.

Iterations continue until all jobs are bottlenecked. To make this procedure
more concrete, consider an example with 4 identical jobs: job 1 with a weight of 3.0,
and jobs 2-4 with a weight of 1.0; and 4 identical GPUs. In the first iteration,
job 1 is assigned resources such that its throughput is 1.0, and jobs 2, 3, and 4 are assigned
resources such that their throughput is 0.33 to respect weights. Job 1 is a bottleneck; the
throughput of the remaining jobs can still be increased. In the next iteration, jobs 2-4
are given full-GPU allocations.

The final allocation satisfies both inter-entity and intra-entity policies.
We note that the above water-filling procedure can also be used for single-level
fairness policies such as the one described in \S\ref{sec:max_min_fairness} to
improve the throughput of non-bottelenecked jobs.

\subsection{Properties of \system{}'s Policies}

Existing scheduling schemes have been analyzed in terms of properties
like sharing incentive, Pareto efficiency, and strategy proofness~\cite{ghodsi2011dominant}.
We formalize \system{}'s heterogeneity-aware policies in the context of
these properties as well.

\paragraph{Homogeneous Clusters.} For homogeneous clusters, \system{}'s heterogeneity-aware policies are
  equivalent to the baseline policies ($\text{throughput}(m, X) = X_m \cdot T_m$), since the heterogeneity-aware
  optimization problems reduce to the original optimization problems
  with one accelerator type.
\paragraph{Sharing Incentive.} For heterogeneous clusters, the policy's objective metric (maximize least
  job share in LAS, completion time of first job in FIFO, or makespan) is at
  least as well off as it would be under a policy that na\"ively
  splits all resources equally among all runnable jobs. This is because the allocation
  corresponding to giving each user $1/n$ of each resource
  is a feasible solution to \system{}'s optimization problem, so
  \system{}'s solution will be at least as good. All \system{} policies
  have \emph{sharing incentive}~\cite{ghodsi2011dominant}, which encourages
  users to use the shared cluster rather than a static private share.
\paragraph{Colocation.} Solutions with colocation
  are always at least as good as without colocation.
\paragraph{Pareto Efficiency.} Allocations of max-min fairness policies
with water filling are Pareto efficient: that is, the allocation for a particular
job cannot be increased without decreasing the allocation for another job.

Note that some of \system{}'s policies may not satisfy other desirable
properties. For example, Sun et al.~\cite{sun2019fair} showed that no
fair-sharing policy can simultaneously satisfy Pareto efficiency, sharing incentive and
strategy proofness in a setting with interchangeable resources. We leave consideration of strategy-proof
policies to future work.

\section{Scheduling Mechanism}
\label{sec:scheduler}

Gavel's scheduling mechanism schedules training iterations of runnable jobs
on the available workers (with possibly different accelerators), such that for each
schedulable job (or combination), the fraction of wall-clock time it spends on
each device class is approximately equal to the computed optimal allocation
$X^{\text{opt}}$ \emph{between} allocation recomputation events. This is challenging for two main
reasons: 1) Jobs can run on multiple accelerators. Moreover, since distributed training can
be communication intensive~\cite{coleman2019analysis, narayanan2019pipedream},
jobs should be placed on accelerators ``close'' to each other (for example,
on accelerators on the same server, or on accelerators in servers in the same  rack).
2) Combinations of up to two jobs can run on a set of accelerators in order to
improve resource utilization (space sharing). Each distinct job
can have $\leq 1$ job combination running in a given round to prevent
work duplication.

\system{} makes its scheduling decisions in \emph{rounds}. That is, the
scheduler tries to place work on all available workers for a specific duration
(this time period is configurable; we use 6 minutes in our experiments).
We call the work handed to each worker in a given round a \emph{micro-task}.
Without rounds, jobs that request many accelerators can suffer from starvation. For
example, consider a cluster with 8 total accelerators and 4 available. The
scheduler can handle a 8-accelerator job waiting for resources in one of two ways:
a) wait for 8 accelerators to become available; 4 accelerators will be unused until the full
quota of 8 accelerators becomes available, b) keep the 8-accelerator job in the queue, and give
4 accelerators to another job that requests a fewer number of resources. However,
this situation can repeat itself, leading to starvation~\cite{zaharia2010delay}.
Scheduling is thus performed in rounds to limit resource under-utilization that
can arise from starvation mitigation, simplify scheduling logic, and still
ensure that jobs that request a large number of workers do not experience
prolonged starvation.

Since the number of active, \emph{schedulable} jobs might far exceed the total
number of workers, \system{} first determines the job combinations that should
run in the upcoming round. To do this, \system{} maintains the time $t_{mj}$ spent by a
job (or combination) $m$ on accelerator type $j$, which is updated as jobs run
on different accelerator types every round. Given $t_{mj}$, \system{}'s scheduler
can then compute the fraction of total wall-clock time spent by
each job (or combination) $m$ on each accelerator type $j$ as $f_{mj} = t_{mj} / (\sum_{m'}t_{m'j})$.
The matrix of priorities is then just the element-wise division of
$X^\text{opt}$ by $f$.

\paragraph{Algorithm.} In every round, we want to move $f_{mj}$ closer to
$X^\text{opt}_{mj}$. This can be achieved by giving high-priority jobs time
on accelerator type $j$.

This problem can be solved exactly if jobs only request single accelerators and if
space sharing is not deployed, by finding the $\text{num\_workers}_j$ jobs with
highest priority (for example, using a heap). However, jobs submitted to
\system{} can be distributed, and space sharing can be used to improve resource
utilization. Solving this problem exactly with these added requirements makes
the problem similar to a multiple-choice knapsack
problem~\cite{sinha1979multiple}, which is NP-hard.

\begin{figure}
    \centering
    \includegraphics[width=1.0\columnwidth]{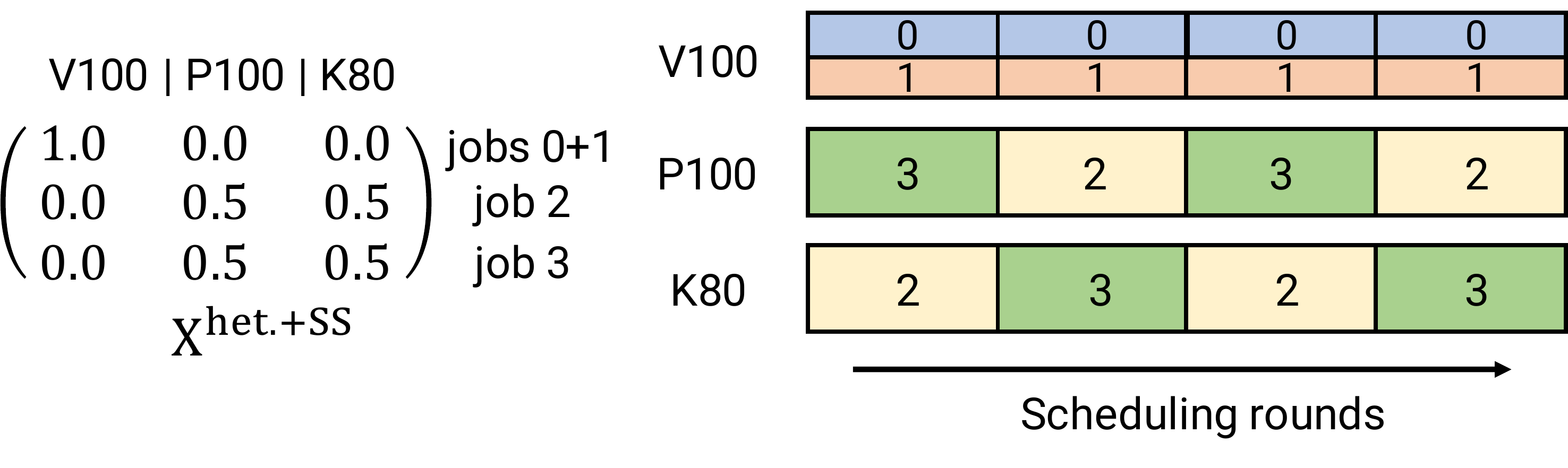}
    \caption{
        \label{fig:scheduling_mechanism}
        Round-based scheduling mechanism in action to
        achieve an allocation $X^\text{het.+SS}$.
        Space sharing is shown with vertically split boxes. Each round
        is denoted by a box. Numbers in boxes represent Job IDs.
    }
    \vspace{-2mm}
\end{figure}

To overcome these challenges, we observe that it is acceptable to make greedy
sub-optimal decisions occasionally, since we can recover from these sub-optimal
decisions in subsequent rounds. We study the impact of this design choice in
\S\ref{sec:evaluation_mechanism}. A job not run in a particular round will have
increased priority in subsequent rounds until it receives accelerator time, while a job
combination that runs in a particular round will have decreased priority. This
ensures that job combinations cannot suffer from starvation if they have a
non-zero optimal allocation (priority $> 0$).

\begin{algorithm}[!t]
\small
\caption{Algorithm for Gavel's scheduling mechanism}
\begin{algorithmic}[1]
\Function{schedule\_jobs}{}
  \State \texttt{active\_combinations} $\leftarrow$ all active job combinations
  \State \texttt{num\_workers\_remaining} $\leftarrow$ number of total workers
  \While{\texttt{num\_workers\_remaining} $> 0$}
    \State $j$ $\leftarrow$ job combination with highest priority
    \State Remove $j$ from \texttt{active\_combinations}
    \If{$j$.\texttt{scale\_factor} $>$ \texttt{num\_workers\_remaining}}
        \State \textbf{continue}
    \EndIf
    \ForAll{$j'$ that conflict (share a job $k$) with $j$}
        \State Remove $j'$ from \texttt{active\_combinations}
    \EndFor
    \State \texttt{num\_workers\_remaining} $-= j$.\texttt{scale\_factor}
  \EndWhile
  \EndFunction
\end{algorithmic}
\label{alg:scheduling_mechanism}
\end{algorithm}

Gavel uses a greedy algorithm to pick the highest-priority job combinations that
fit in the provided resource budget. The algorithm maintains a set of eligible
job combinations (\texttt{eligible\_job\_combinations}) that can be
scheduled in the upcoming scheduling round. The scheduling mechanism then tries
to add job combinations with highest priority into a
\texttt{job\_combinations\_to\_schedule} set. Once a job combination is added
to this set, all \emph{conflicting} job combinations are removed from the set
of eligible combinations to ensure that a given job is not run more than once
in a given scheduling round. Job combinations that cannot fit in the current
round due to space limitations (required number of accelerators $\text{scale\_factor}$ unavailable) are also
removed from the set of eligible combinations. This algorithm is
detailed in Algorithm~\ref{alg:scheduling_mechanism}. \system{}'s scheduling
mechanism is decoupled from its policies, ensuring that the same scheduling
mechanism can be used for many different policies.
Figure~\ref{fig:scheduling_mechanism} shows \system{}'s scheduling mechanism in
action.

Once \system{} has decided what jobs (and combinations) should run in a given
round on different accelerator types, \system{} must decide how to \emph{place}
these jobs. \system{}'s scheduler places jobs in decreasing order of the number
of requested workers, and tries to give jobs accelerators on the same physical
server to minimize fragmentation.

\section{Implementation} \label{sec:implementation}

We implemented a prototype of Gavel and an accompanying simulator in approximately
8000 lines of Python code. We used \texttt{cvxpy}~\cite{diamond2016cvxpy} to
implement \system{}'s heterogeneity-aware policies, and \texttt{gRPC}~\cite{grpc} for
the communication of control messages between the scheduler and workers.

\paragraph{Interface between Scheduler and Applications.} \system{} currently
supports user applications written in PyTorch~\cite{pytorch}; support for
TensorFlow~\cite{tensorflow} is left for future work. The scheduler and user applications
then interact through a narrow API. \system{} ships with
a Python library that users can import into their code. This library provides
an implementation for a \texttt{GavelIterator}, a wrapper around existing
framework-provided data iterators. The \texttt{GavelIterator} ensures that each
task in a distributed job runs for the same number of iterations, and synchronizes
the conclusion of rounds between the scheduler and workers.
\texttt{GavelIterator} is instantiated with \texttt{load\_checkpoint} and
\texttt{save\_checkpoint} function pointers, which can be called to load all
necessary parameters and metadata from a checkpoint at the start of a round, and
to create a checkpoint at the end of a round. Both the \texttt{load\_checkpoint}
and \texttt{save\_checkpoint} methods need to be implemented by the user,
but only need to call appropriate framework
methods (about 10 LOC in our implementation).
\texttt{GavelIterator} contacts the scheduler near the end of a round to see
if the same job will run in the next round on the same worker. We call this a
\emph{lease renewal}. If the lease is not renewed, the iterator calls the
\texttt{save\_checkpoint} method at the end of the round, and returns execution
back to the scheduler. The scheduler can then launch another job on the worker.

\paragraph{Throughput Estimation.}

\begin{figure}[t]
  \includegraphics[width=1.0\linewidth]{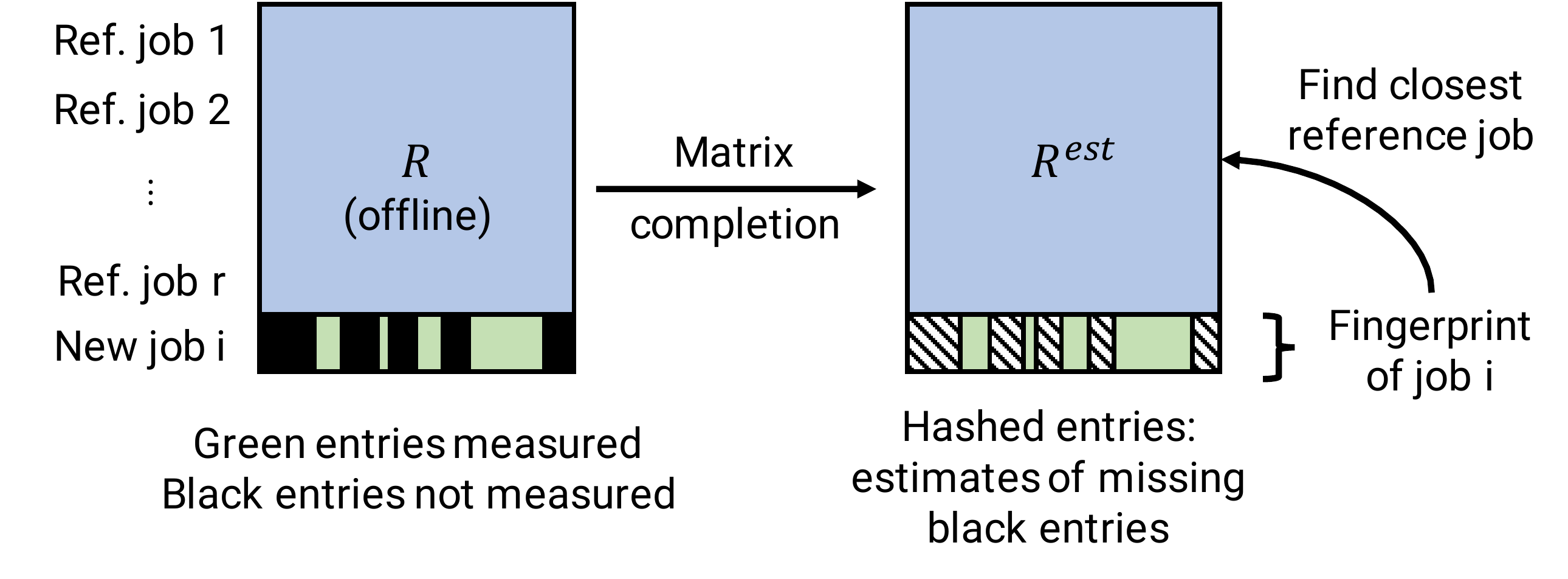}
  \vspace{-0.25in}
  \caption{\system{}'s throughput estimator. Profiling is combined with matrix
  completion to obtain a fingerprint for every new job. The fingerprint is then
  used to find the closest reference job.}
  \label{fig:throughput_estimation_procedure}
\end{figure}

Gavel uses a similar technique to Quasar~\cite{delimitrou2014quasar}
to estimate colocated throughputs when space sharing if they are not available a priori, mixing profiling
with matrix completion. Matrix completion enables sparse low rank matrices to
be reconstructed with low error~\cite{mnih2008probabilistic,candes2010matrix}.
With matrix completion, Gavel is able to extrapolate measurements obtained
through direct profiling on separate workers dedicated to profiling, and determine the job's most similar pre-profiled
reference job. The throughput estimator can then use the reference job's
throughput measurements as an initial throughput estimate. \system{}'s
throughput estimator is diagrammed in Figure~\ref{fig:throughput_estimation_procedure}.

\begin{table}[t!]
\resizebox{\columnwidth}{!}{
\begin{tabular}{llll}
\toprule
\textbf{Model}                                                           & \textbf{Task}                                                                 & \begin{tabular}[l]{@{}c@{}}\textbf{Dataset /}\\ \textbf{Application}\end{tabular} & \textbf{Batch size(s)}                                                           \\
\toprule
ResNet-50~\cite{he2016deep, resnet50}                                                       & \begin{tabular}[l]{@{}l@{}}Image\\ Classification\end{tabular}       & ImageNet~\cite{deng2009imagenet}                                                        & \begin{tabular}[l]{@{}l@{}}16, 32,\\ 64, 128\end{tabular}          \\ \hline
ResNet-18~\cite{he2016deep, resnet18}                                                       & \begin{tabular}[l]{@{}l@{}}Image\\ Classification\end{tabular}       & CIFAR-10~\cite{krizhevsky2014cifar}                                                        & \begin{tabular}[l]{@{}l@{}}16, 32, 64,\\ 128, 256\end{tabular}          \\ \hline
A3C~\cite{mnih2016asynchronous, a3c}                                                             & Deep RL                                                              & Pong                                                            & 4                                                                       \\ \hline
LSTM~\cite{lm}                                                            & \begin{tabular}[l]{@{}l@{}}Language\\ Modeling\end{tabular}          & Wikitext-2~\cite{merity2016pointer}                                                      & \begin{tabular}[l]{@{}l@{}}5, 10, 20,\\ 40, 80\end{tabular}             \\ \hline
Transformer~\cite{vaswani2017attention, transformer}                                                     & \begin{tabular}[l]{@{}l@{}}Language\\ Translation\end{tabular}       & \begin{tabular}[l]{@{}l@{}}Multi30k~\cite{W16-3210}\\ (de-en)\end{tabular}      & \begin{tabular}[l]{@{}l@{}}16, 32, 64,\\ 128, 256\end{tabular}          \\ \hline
CycleGAN~\cite{zhu2017unpaired, cyclegan}                                                        & \begin{tabular}[l]{@{}l@{}}Image-to-Image\\ Translation\end{tabular} & monet2photo~\cite{zhu2017unpaired}                                                     & 1                                                                       \\ \hline
\begin{tabular}[l]{@{}l@{}}Recoder~\cite{abdallah2018recoder}\\ (Autoencoder)\end{tabular} & Recommendation                                                       & ML-20M~\cite{harper2016movielens}                                                          & \begin{tabular}[l]{@{}l@{}}512, 1024,\\ 2048, 4096,\\ 8192\end{tabular} \\ \bottomrule
\end{tabular}}
\vspace{-0.1in}
\caption{Models used in evaluation.}
\vspace{-0.1in}
\label{table:model_list}
\end{table}

\begin{table}[t!]
\footnotesize
\centering
\begin{tabular}{lllll}
\toprule
\textbf{Trace} & \textbf{System} & \textbf{Objective} & \textbf{Physical} & \textbf{Simulation} \\
\toprule
Continuous & Gavel  & Average JCT & 3.6 hrs & 3.4 hrs \\
Continuous & Baseline LAS       & Average JCT & 5.0 hrs & 5.2 hrs \\
\midrule
Static & Gavel & Makespan & 17.7 hrs & 17.9 hrs \\
Static & Gandiva       & Makespan & 21.3 hrs & 21.9 hrs \\ \bottomrule
\end{tabular}
\caption{Comparison of end objective between physical experiment and simulation
         for two different traces. For the continuous trace, we measure the average JCT
         of 25 jobs in a steady-state cluster. For the static trace, we
         measure the total time needed to complete 100 jobs submitted
         at the start of the run. The heterogeneity-aware policies improve target
         objectives by up to 1.4$\times$, and results on the physical cluster
         are in agreement with results on simulated cluster ($<5\%$).}
\label{table:physical_vs_simulation}
\vspace{-0.1in}
\end{table}

\begin{figure}
    \center
    \begin{subfigure}[b]{\columnwidth}
        \includegraphics[width=0.85\columnwidth]{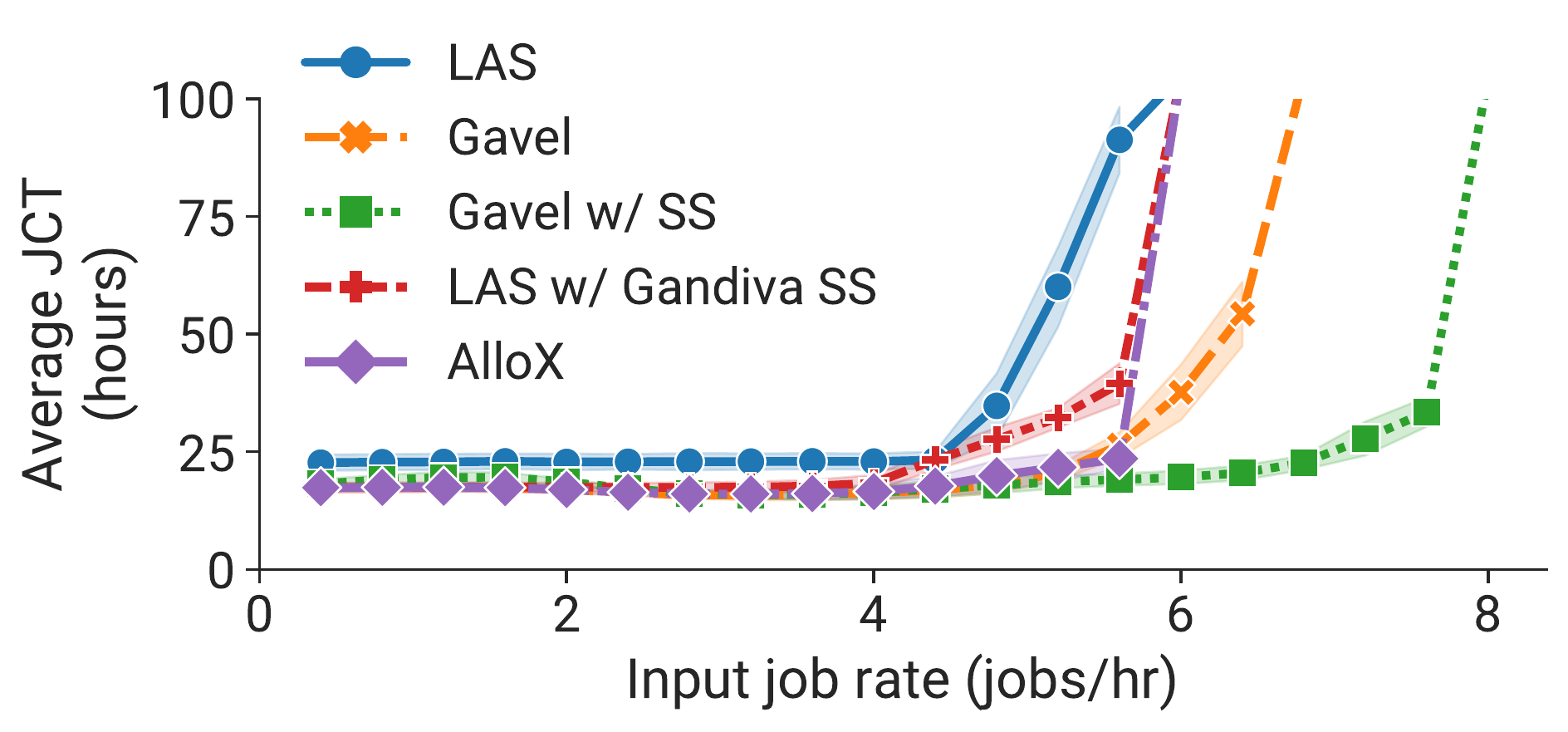}
        \caption{Average job completion time vs. cluster load.}
    \end{subfigure}
    \begin{subfigure}[b]{\columnwidth}
        \includegraphics[width=0.85\columnwidth]{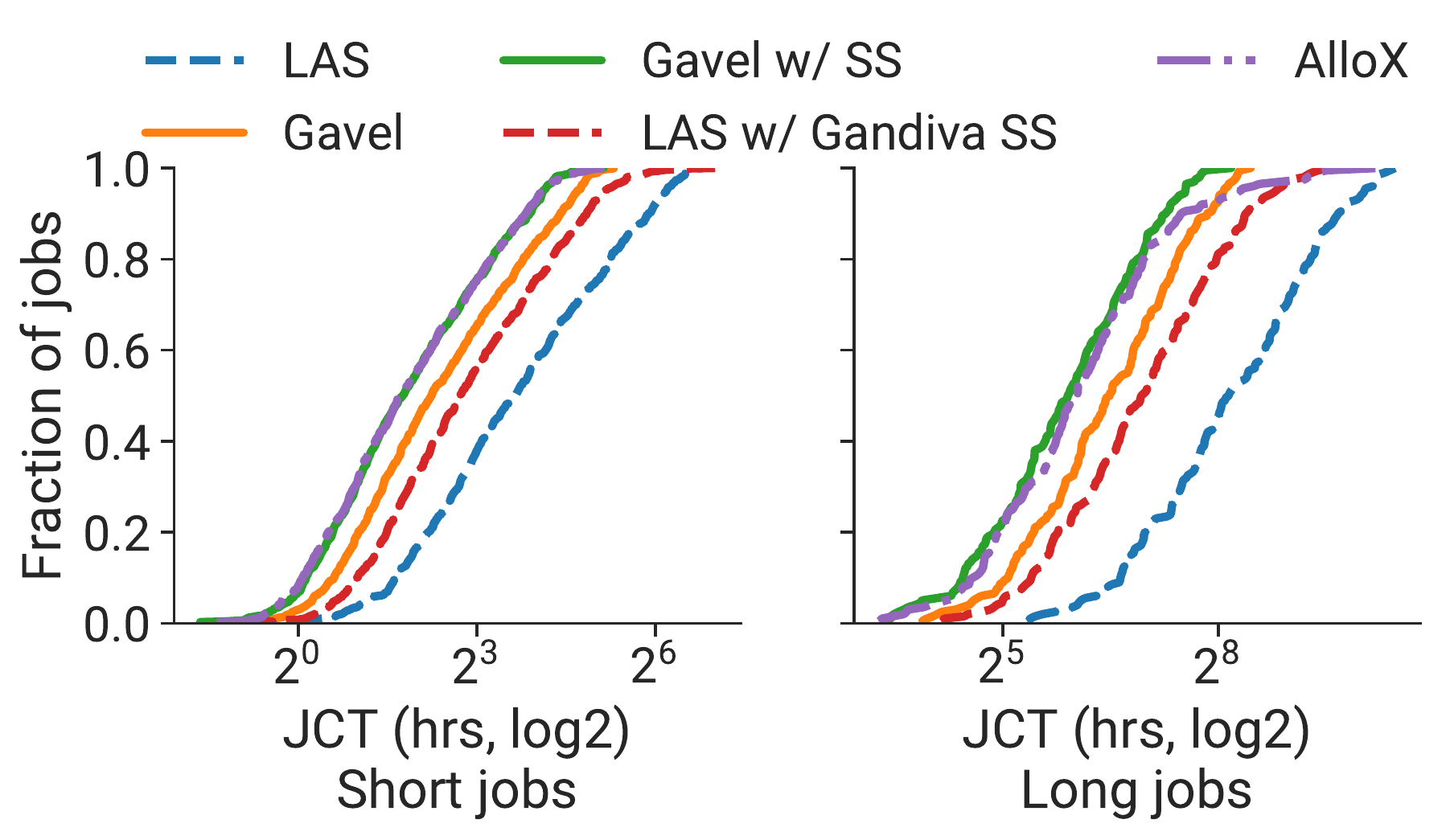}
        \caption{CDF of job completion times.}
    \end{subfigure}
    \caption{
        Comparison of heterogeneity-agnostic least attained service (LAS) policy
        to a heterogeneity-aware LAS policy, on the \textit{simulated cluster and
        continuous-single trace}.
        \label{fig:las_single_gpu}
    }
    \vspace{-0.1in}
\end{figure}

\section{Evaluation} \label{section:evaluation}

In this section, we seek to answer the following questions:
\begin{itemize}
\item Do \system{}'s heterogeneity-aware policies improve objective metrics in
    a physical cluster (\S\ref{sec:evaluation_physical}) and in simulations of
    larger clusters (\S\ref{sec:evaluation_simulation})?
\item How does \system{}'s scheduling overhead scale with the number of active jobs?
   (\S\ref{sec:evaluation_scalability})
\item How well does \system{}'s scheduling mechanism realize \system{}'s heterogeneity-aware allocations?
   (\S\ref{sec:evaluation_mechanism})
\item Is \system{} able to accurately estimate the throughputs of co-located jobs
  when using space sharing? (\S\ref{sec:eval_throughput_estimation})
\end{itemize}

\subsection{Experiment Setup}

We run experiments on both a physical and simulated cluster.

\paragraph{Clusters.} We run physical cluster experiments on a 48-GPU cluster
with 8 V100s, 16 P100s, and 24 K80s. Simulated cluster experiments are run on
a 108-GPU cluster with 36 V100s, 36 P100s, and 36 K80s.

\paragraph{Traces.} We run physical and simulated experiments on two types of
traces: one where all jobs are available at the start of the trace and
jobs are \emph{not} subsequently added (``static''), and another where jobs are continuously
added to the cluster (``continuous''). 

For the continuous trace, job arrival
times are generated according to a Poisson arrival process initialized with an
inter-arrival rate $\lambda$. For the simulated experiments, we vary $\lambda$
to show the extra load each \system{} policy is able to sustain in steady state.
We run 3 seeds for every $\lambda$, and show standard deviations.
For the physical cluster experiments, we use a single $\lambda$ that keeps the
cluster well-utilized in steady state. 

Traces are populated with a list of 26
job configurations, including CNN-based image classification models,
RNN-based language models, and others, based on performance measurements on real hardware.
Table~\ref{table:model_list} presents the full list of models and configurations.
We sample durations from an exponential distribution between $10^{1.5}$ minutes
and $10^{4}$ minutes to match the process Gandiva used in its
evaluation~\cite{xiao2018gandiva}. For the simulated experiments, we show
results in two regimes: one where all jobs use a single worker (``continuous-single''),
and another where roughly 70\% of jobs request a single worker, another 25\%
request between 2 and 4 workers, and the remaining 5\% request 8 workers,
as observed in published traces from Microsoft~\cite{jeon2019github} (``continuous-multiple'').

\paragraph{Metrics.} For fairness and FIFO policies, our target metric is average
job completion time of steady-state jobs, which is the same metric used by related
work~\cite{mao2019learning, gu2019tiresias}. We also show finish time fairness (FTF)
for policies that explicitly optimize for FTF. For makespan policies, our target
metric is the time needed to complete a batch of jobs, and only
the static trace is used. For cost-related
policies, the metric is cost (in dollars), and the percentage of jobs that violate
SLOs.

\paragraph{Round durations.} For the results presented in \S\ref{sec:evaluation_physical},
we use a round duration of 20 minutes, and for all other results we use a round duration
of 6 minutes (we explain the choice of 6 minutes in \S\ref{sec:evaluation_mechanism}).
We use a larger duration in the physical cluster experiments to mitigate
scheduler overhead; however, previous work has shown that it is feasible to hide
this overhead~\cite{xiao2018gandiva, mahajan2020themis, elasticdl, chen2017preemptive}.
We also found that using 6 minute rounds in simulation for the \S\ref{sec:evaluation_physical}
experiments did not significantly alter the results.
\begin{figure}
    \center
    \begin{subfigure}[b]{\columnwidth}
        \includegraphics[width=0.85\columnwidth]{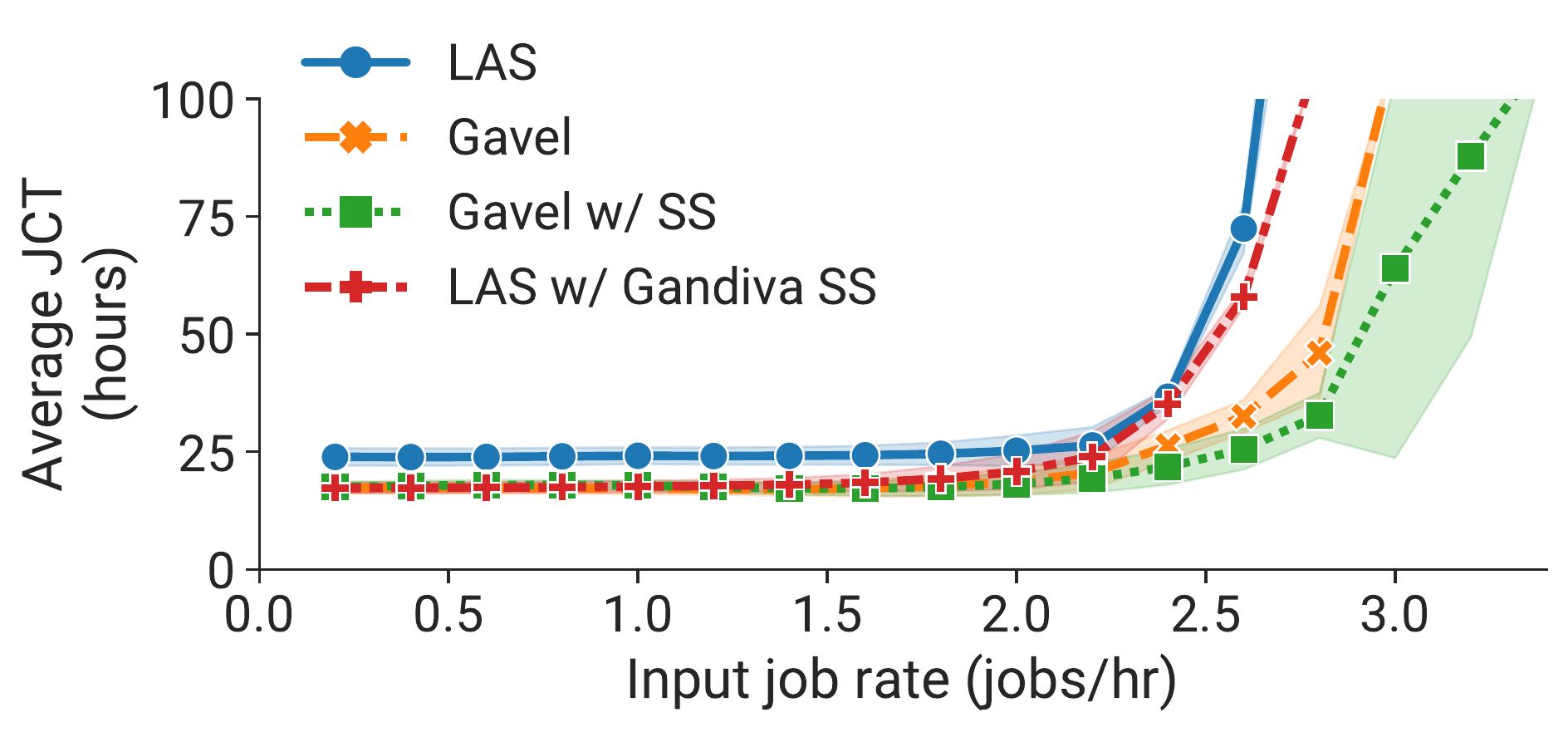}
        \caption{Average job completion time vs. cluster load.}
    \end{subfigure}
    \begin{subfigure}[b]{\columnwidth}
        \includegraphics[width=0.85\columnwidth]{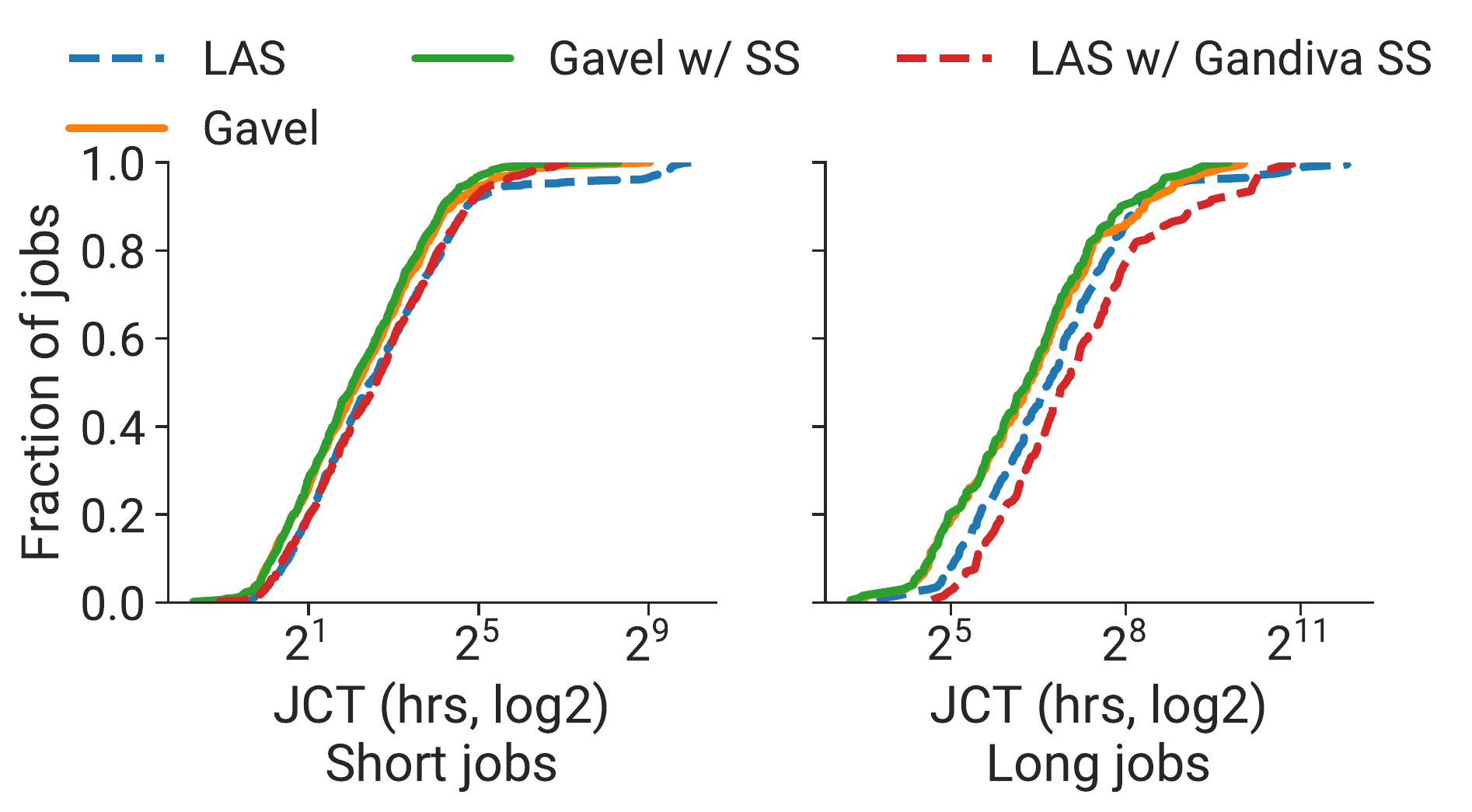}
        \caption{CDF of job completion times.}
    \end{subfigure}
    \caption{
        Comparison of heterogeneity-agnostic least attained service (LAS) policy
        to a heterogeneity-aware LAS policy, on the simulated cluster and
        continuous-multiple trace. Each input job rate is run with 3 seeds; shaded
        regions around lines show the standard deviati/n.
        \label{fig:las_multi_gpu}
    }
    \vspace{-0.1in}
\end{figure}

\subsection{End-to-End Results on Physical Cluster}
\label{sec:evaluation_physical}

For our physical cluster experiments, we run a heterogeneity-aware and a
heterogeneity-agnostic fairness policy on an
appropriately-sized continuous trace, and a heterogeneity-aware makespan policy against a baseline
that uses Gandiva's ad-hoc space sharing on an appropriately-sized static trace.
Results are shown in Table~\ref{table:physical_vs_simulation}. \system{}'s
heterogeneity-aware policies help improve objectives by up to
\textbf{1.4$\times$}.

We also compare the real performance to simulations and observe that
for both policies, the difference between metrics in
simulation and on the physical cluster is small ($<5\%$), indicating that our
simulator has high fidelity.

\subsection{End-to-End Results in Simulation} \label{sec:evaluation_simulation}

We use a larger simulated cluster to evaluate the efficacy of \system{}'s
heterogeneity-aware policies across a range of objectives, and compare with
heterogeneity-agnostic versions from previous work. As appropriate, we compare to other baselines like AlloX.
 We include additional supporting figures in the Appendix (\S\ref{app:sec-additional_results}).

\paragraph{Least Attained Service (LAS).} Figures~\ref{fig:las_single_gpu} and~\ref{fig:las_multi_gpu}
compare the vanilla LAS policy with its heterogeneity-aware variants.
We compare with two other baselines: a modified LAS policy that uses
Gandiva's ad-hoc space sharing, and an AlloX policy that explicitly optimizes
average job completion time (but only for single-worker jobs).
We make three observations.

First, the heterogeneity-aware policies
support higher load on the \emph{same} cluster, and also reduce
average JCT by up to \textbf{3.5$\times$} for the single-worker
trace, and by up to \textbf{2.2$\times$} for the multi-worker trace.
Second, the heterogeneity-aware LAS version supports higher load
than AlloX, since AlloX can give short jobs preferential treatment in the
interest of optimizing average JCT, leading to long jobs experiencing
starvation (long tail in long jobs CDF). At moderate load, AlloX represents a best-case
scenario since it explicitly optimizes for average JCT on a heterogeneous
cluster. \system{} is able to essentially match AlloX on average JCT, while
also being able to support other objectives.
Third, Gandiva-style packing, which randomly explores job combinations
until a combination that improves performance is found, is ineffective compared
to the principled packing used by \system{}'s heterogeneity-aware policies
(\textbf{2.2$\times$} better average JCT for both traces).

\begin{figure}
    \center
    \begin{subfigure}[b]{\columnwidth}
        \includegraphics[width=0.85\columnwidth]{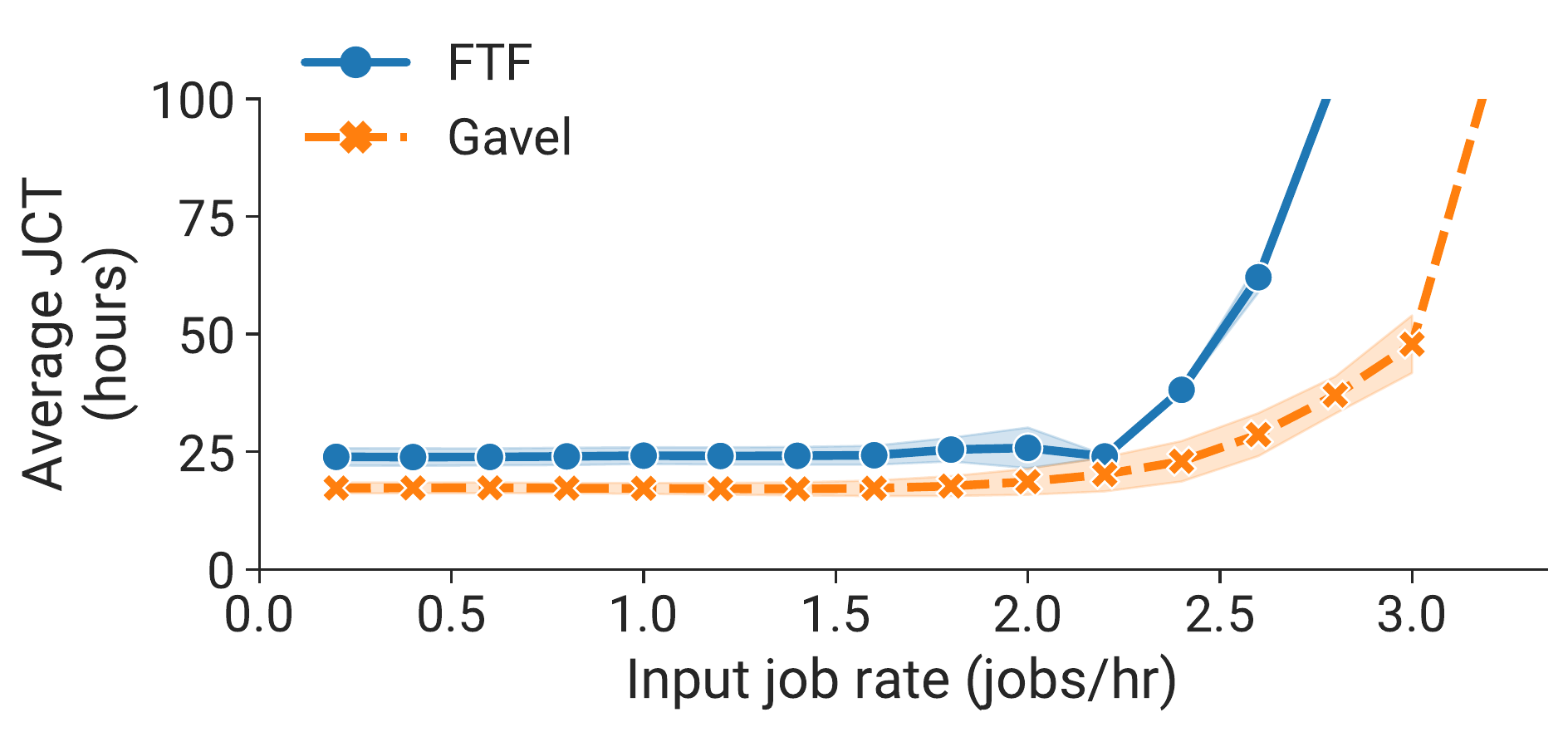}
        \caption{Average job completion time vs. cluster load.}
    \end{subfigure}
    \begin{subfigure}[b]{\columnwidth}
        \includegraphics[width=0.85\columnwidth]{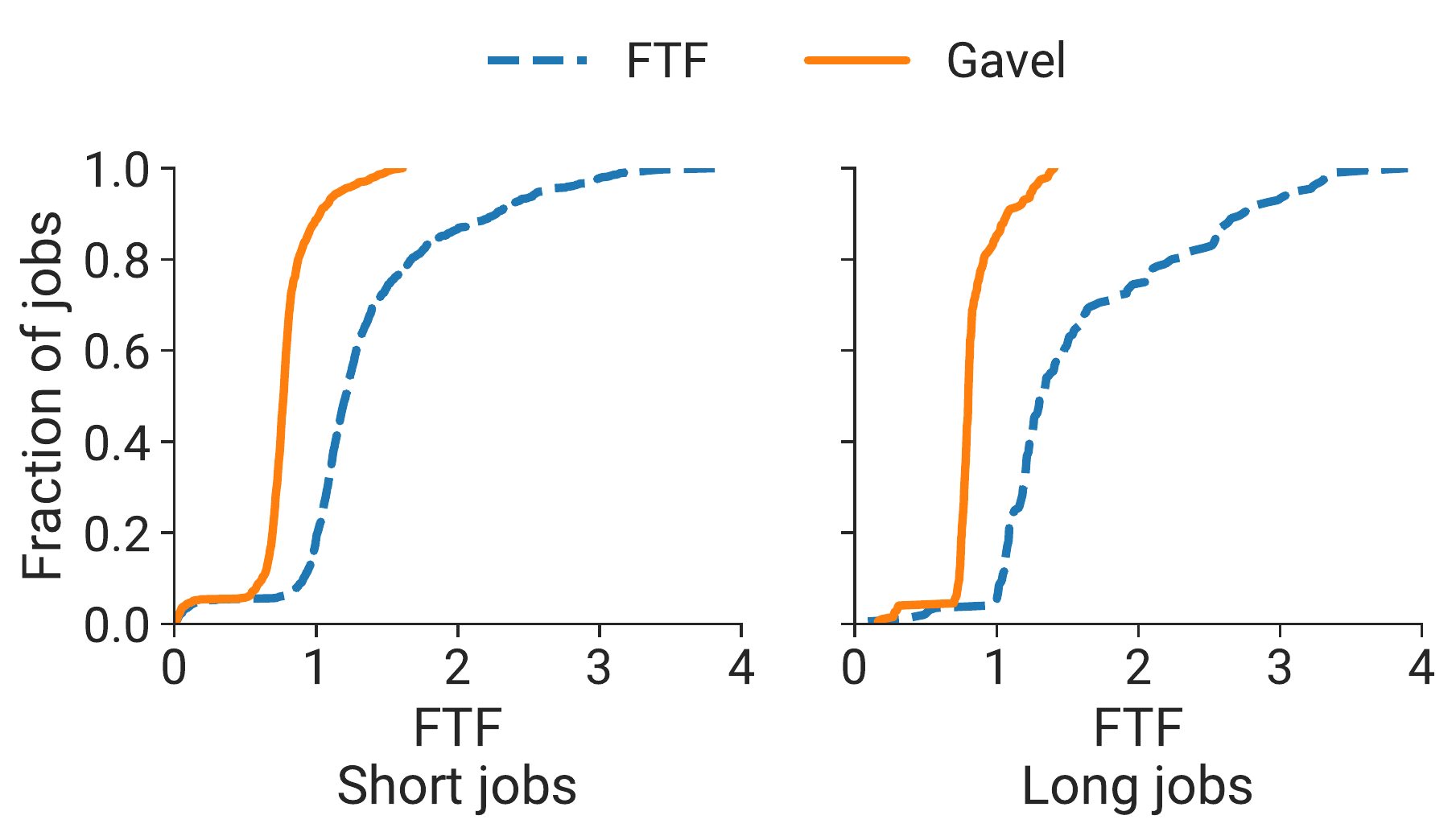}
        \caption{CDF of perf-job FTF.}
    \end{subfigure}
    \caption{
        Comparison of heterogeneity-agnostic finish time fairness (FTF) policy
        to a heterogeneity-aware FTF policy, on the simulated cluster and
        continuous-multiple trace.
        \label{fig:ftf_multi_gpu}
    }
    \vspace{-0.1in}
\end{figure}

\paragraph{Finish Time Fairness (FTF).} We compare the heterogeneity-aware version
of Finish Time Fairness (FTF) to its heterogeneity-agnostic counterpart in
Figure~\ref{fig:ftf_multi_gpu}. The heterogeneity-aware policy reduces average JCTs 
by \textbf{3$\times$} and improves average FTF by \textbf{2.8$\times$}.

\paragraph{Makespan.} \system{}'s heterogeneity-aware makespan policy reduces makespan by \textbf{2.5$\times$}
compared to a FIFO baseline, and by \textbf{1.4$\times$} compared to a baseline that
uses Gandiva's ad-hoc space sharing. Makespan is reduced by a further \textbf{8\%}
when the number of jobs in the trace is high when using space sharing.

\paragraph{FIFO.}
The heterogeneity-aware versions of FIFO allow the cluster to support higher load (in terms
of average input job rate). At high load, the heterogeneity-aware version
without space sharing reduces average JCT by up to \textbf{2.7$\times$}, and the
heterogeneity-aware version with space sharing reduces average JCT by up to
\textbf{3.8$\times$}. Space sharing is less effective for distributed jobs: it reduces
average JCT by \textbf{1.1$\times$} for the trace with distributed jobs, and by \textbf{1.4$\times$}
for the single-GPU trace.

\paragraph{LAS with priorities.} We also run an experiment with the LAS policies
where 20\% of jobs have higher priority. At high load, \system{} reduces the
average JCT of high-priority jobs by \textbf{1.5$\times$} and the average JCT
of low-priority jobs by \textbf{2.7$\times$}.

\begin{figure}[t]
    \centering
    \begin{subfigure}[b]{1.0\columnwidth}
        \centering
        \includegraphics[width=0.85\columnwidth]{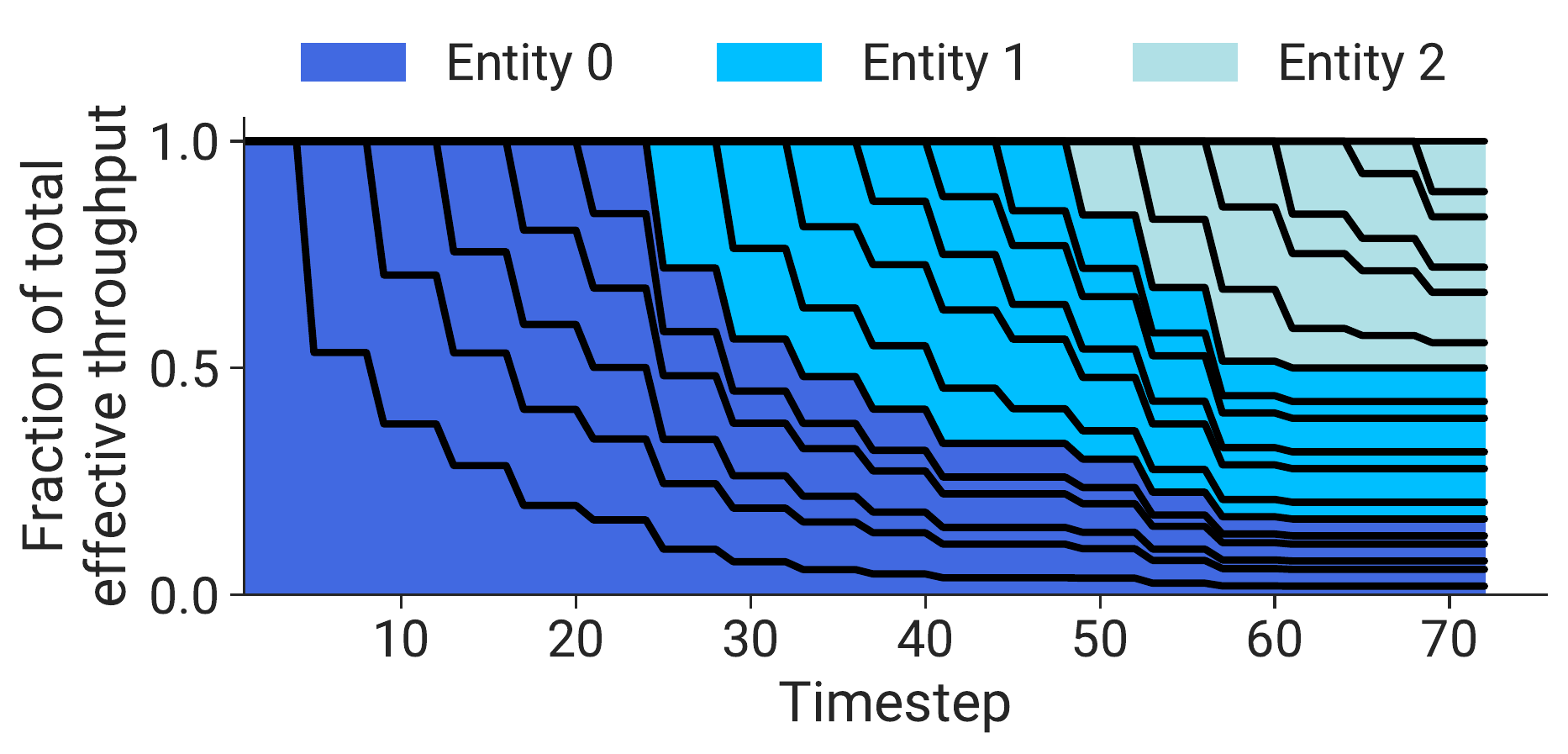}
        \caption{\small Fraction of total throughput for each job with time.}
        \label{fig:hierarchical_timeline_fraction}
    \end{subfigure}
    \begin{subfigure}[b]{1.0\columnwidth}
        \centering
        \includegraphics[width=0.85\columnwidth]{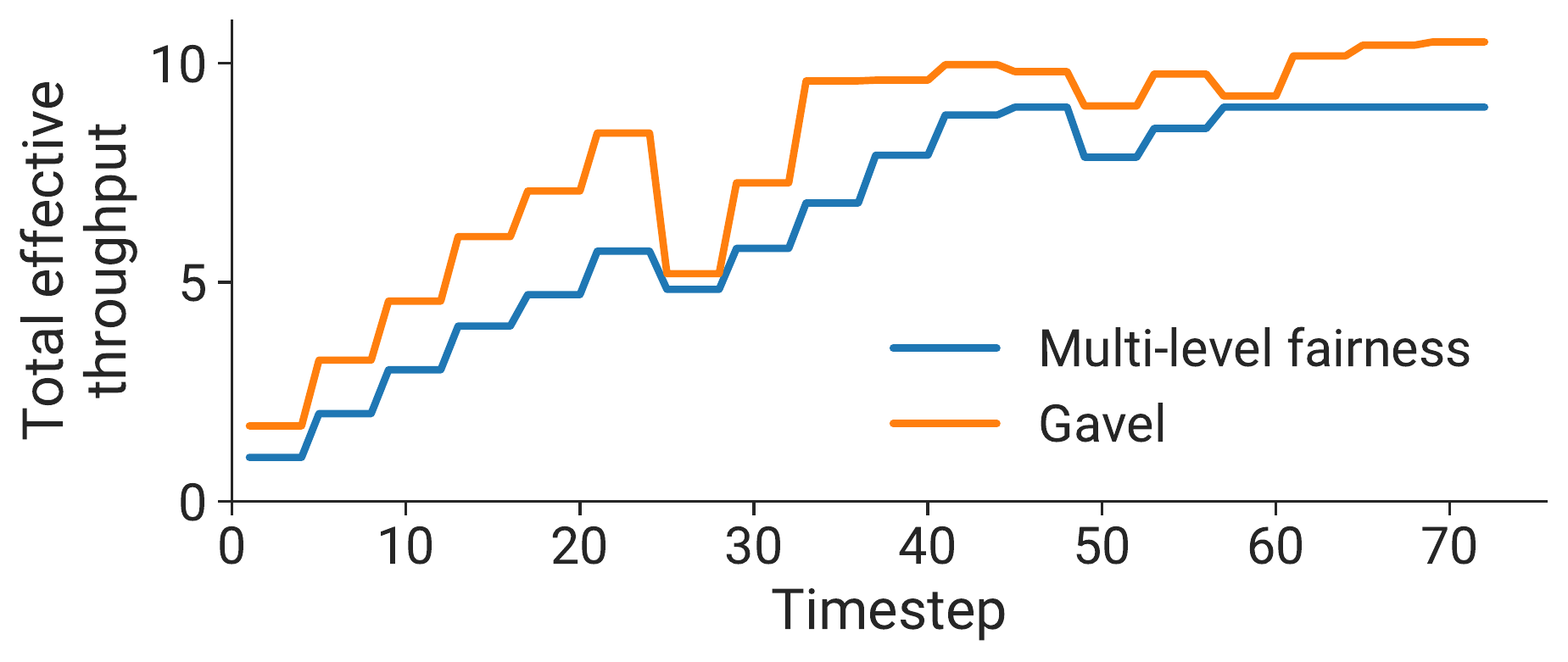}
        \caption{\small Total throughput vs. time.}
        \label{fig:hierarchical_total}
    \end{subfigure}
    \caption{
        \label{fig:hierarchical}
        Behavior of a multi-level fairness policy with time as jobs are added to a
        small cluster with 3 V100 GPUs, 3 P100 GPUs, and 3 K80 GPUs.
        Each line represents a separate job, and jobs are added every 4
        timesteps. The first 6 jobs belong to entity 0 (weight of entity, $w_0 = 1$),
        the next 6 jobs belong to entity 1 ($w_1 = 2$), and the last 6 jobs belong to
        entity 2 ($w_2 = 3$).
    }
    \vspace{-0.1in}
\end{figure}

\paragraph{Cost.}
We simulate each of the cost policies on a 500-job workload comprised of
ResNet-50 and A3C jobs. As we observe in Figure~\ref{fig:heterogeneity-cost},
the ResNet-50 job has the best cost-normalized throughput on the V100 while the
A3C job has the best cost-normalized throughput on the K80. Each job's duration
is chosen from $\{0.5, 1, 2, 4, 8\}$ days, and each job's SLO is chosen from
$\{1.2\times, 2\times, 10\times\}$ its duration.

The policy that minimizes
cost reduces the total cost compared to the policy that maximizes throughput
by a factor of roughly \textbf{1.4$\times$}. However, approximately \textbf{35\%} of jobs violate
their SLO as this policy prioritizes cheaper but slower GPUs; in particular,
the A3C jobs are scheduled on K80 GPUs which results in violations for
tight SLOs. In comparison, the policy that includes SLOs as well eliminates all
violations for a small increase in cost (a cost reduction of \textbf{1.23$\times$}
compared to the baseline policy), by ensuring that A3C jobs with tight SLOs are
run on instances with V100 GPUs.

\paragraph{Multi-level Hierarchical Policies.}
Figure~\ref{fig:hierarchical} shows the behavior of a multi-level fairness policy
as new jobs belonging to multiple entities are added to the cluster. Resources
are granted to jobs in a way that respects both
the higher-level and lower-level policies: in Figure~\ref{fig:hierarchical_timeline_fraction},
fairness is enforced both within and across entities (as can be seen by
the widths of the colored bands, which represents cross-entity fairness,
and the widths of bands within a color, which represents fairness across jobs within an
entity), and allocations are adjusted as new jobs come in. The Figure~\ref{fig:hierarchical_fairness_fifo} in the Appendix shows
a similar timeline diagram for a fairness+FIFO policy.

The multi-level fairness policy can also be implemented in a
heterogeneity-agnostic manner by statically partitioning resources across users
while respecting per-entity and per-user weights. While this results in a
fair allocation as well, we observe that total effective throughput is about
\textbf{17\%} lower compared to the heterogeneity-aware policy (Figure~\ref{fig:hierarchical_total}).

\subsection{Scalability of \system{}} \label{sec:evaluation_scalability}

\begin{figure}[t]
    \centering
    \includegraphics[width=0.6\columnwidth]{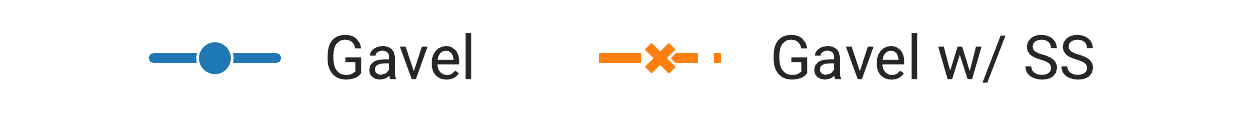}
    \begin{subfigure}[b]{0.48\columnwidth}
        \centering
        \includegraphics[width=1.0\columnwidth]{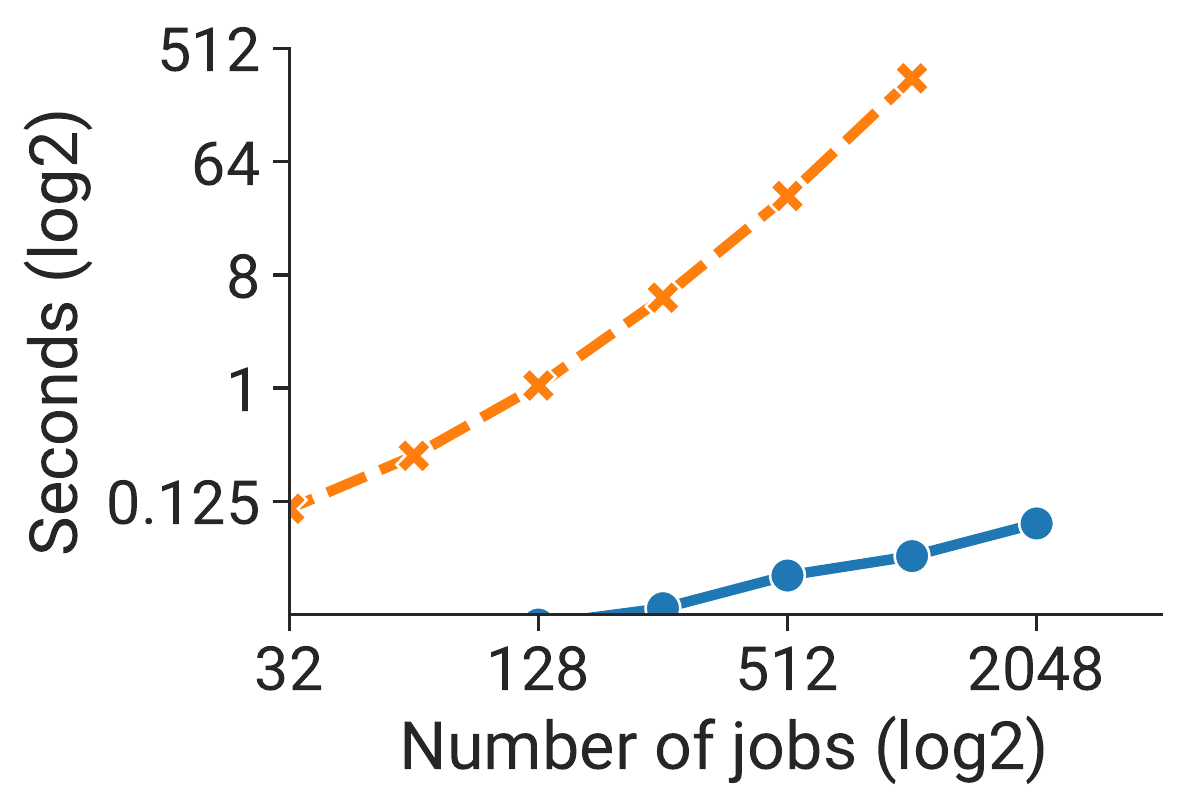}
        \caption{
        \small LAS.}
    \end{subfigure}
    \begin{subfigure}[b]{0.48\columnwidth}
        \centering
        \includegraphics[width=1.0\columnwidth]{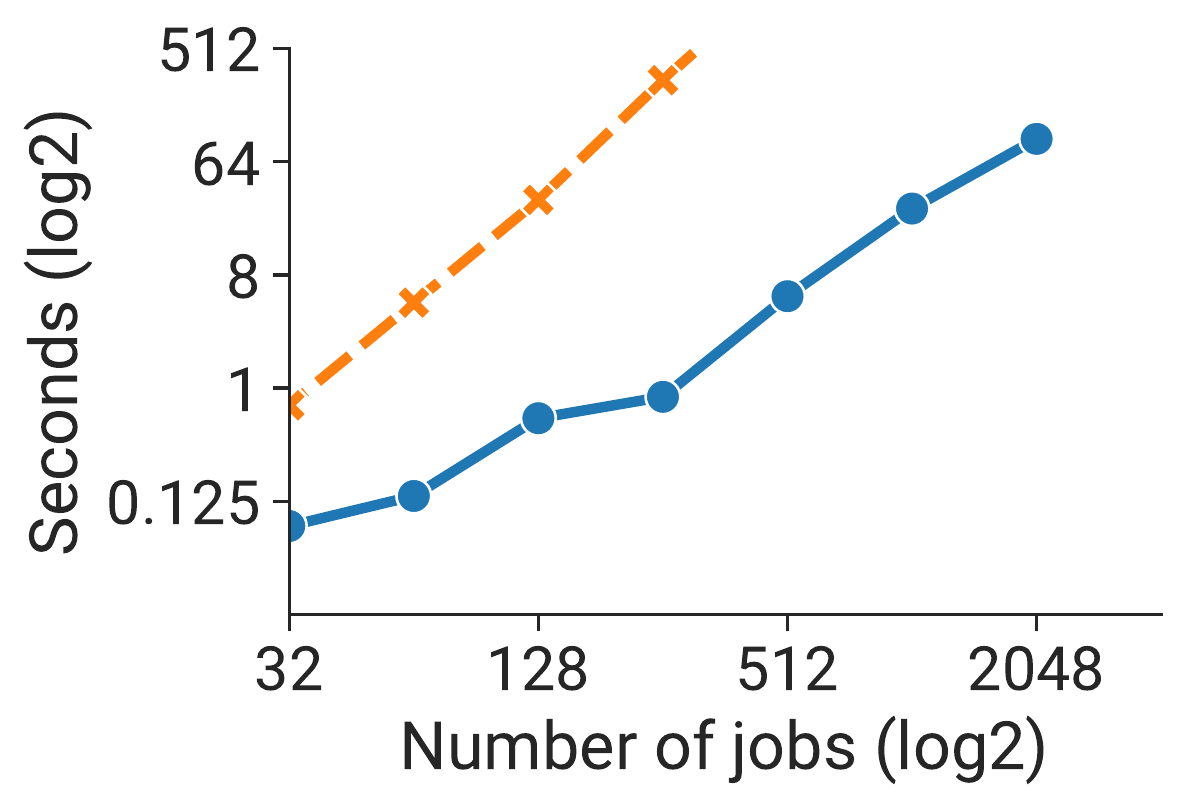}
        \caption{
        \small Hierarchical.}
    \end{subfigure}
    \caption{
        \label{fig:policy_runtimes}
        Scaling of LAS and hierarchical policies with the number of active jobs on
        a heterogeneous cluster with an equal number of V100, P100, and K80 GPUs.
        The size of the cluster is increased as the number of active jobs is
        increased.
    }
\end{figure}

Figure~\ref{fig:policy_runtimes} shows the scaling behavior of the heterogeneity-aware
LAS and multi-level fairness policies with and without space sharing.
We observe that even with 2048 active jobs, the hierarchical policy without space
sharing can be run in $<10$ minutes. With space sharing, the policy can be run
with 512 jobs in $<10$ minutes. The single-level LAS policy is much cheaper
to compute in comparison. We note that allocations do not need to be recomputed
every scheduling round -- however, the longer the policy takes to run, the longer
it takes for a new job to be granted resources on the cluster. Having said that,
we believe latencies of $<30$ minutes are acceptable for large clusters (and preferable
to a non-preemptive scheduler where jobs might experience
large queuing delays).

\subsection{Efficacy of Scheduling Mechanism} \label{sec:evaluation_mechanism}

\begin{figure}[t]
    \centering
    \begin{subfigure}[b]{0.48\columnwidth}
        \centering
        \includegraphics[width=1.0\columnwidth]{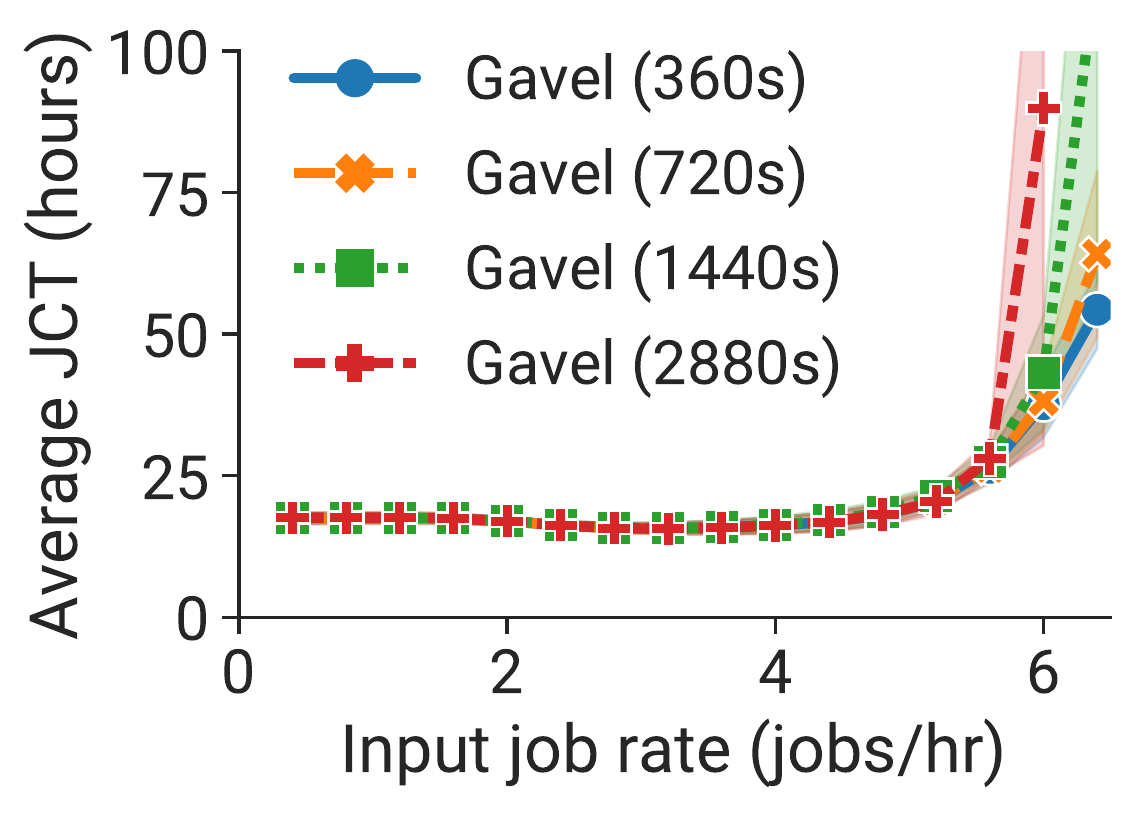}
        \caption{
        \label{fig:time_per_iteration_sweep}
        \small Effect of round length.}
    \end{subfigure}
    \begin{subfigure}[b]{0.48\columnwidth}
        \centering
        \includegraphics[width=1.0\columnwidth]{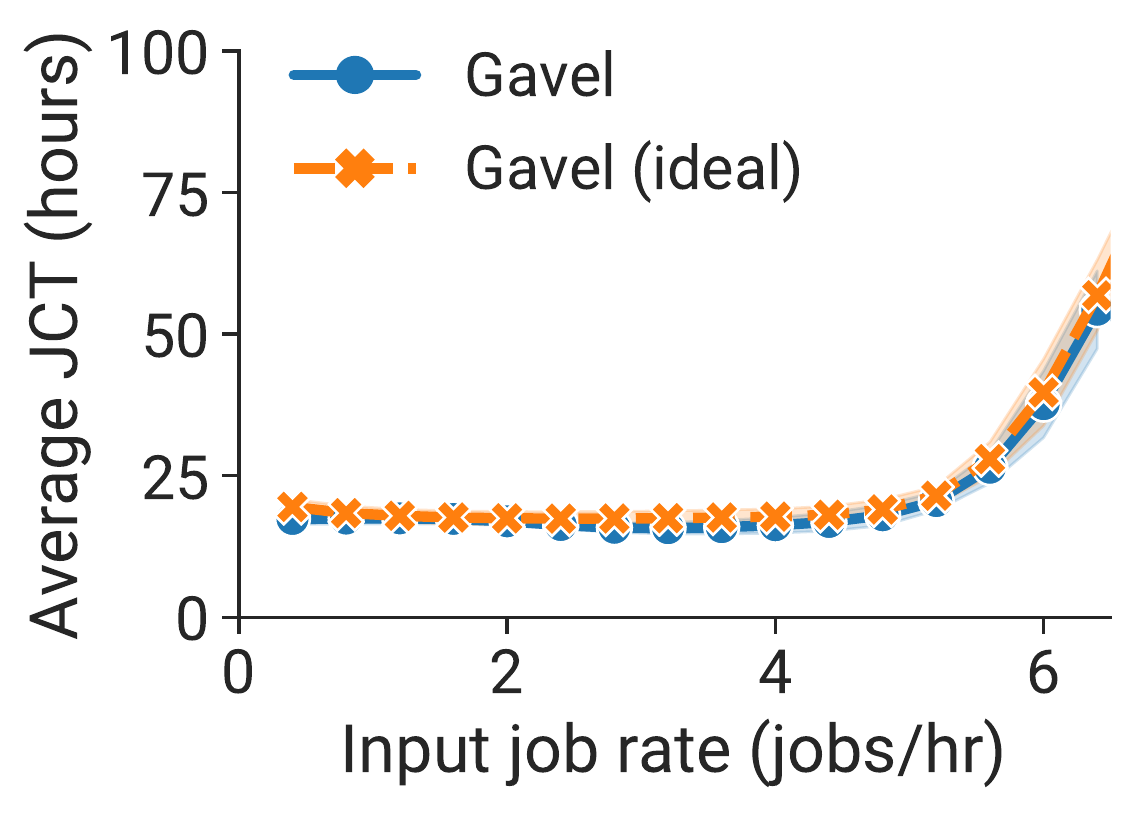}
        \caption{
        \label{fig:mechanism_vs_ideal}
        \small Mechanism vs. ideal.}
    \end{subfigure}
    \caption{
        \label{fig:scheduler_mechanism}
        (a) Effect of round length on average JCT for the heterogeneity-aware LAS
        policy. (b) Comparison of scheduling mechanism to an ideal baseline that allocates
        resources to jobs \emph{exactly} according to the computed allocation
        for the same policy.
    }
    \vspace{-0.1in}
\end{figure}

Figure~\ref{fig:time_per_iteration_sweep} shows the effect of the round length
on average JCT for the heterogeneity-aware LAS policy with a
single-GPU trace. We observed similar behavior on traces with multi-GPU jobs, as well as other policies.
A smaller round length gives \system{}'s scheduling mechanism
more rounds to course correct, allowing the true allocation and computed optimal
allocation to more closely match. We found that the time needed to load and
save checkpoints for our target models is $< 5$ seconds, which means that a
round length of 6 minutes gives a good theoretical tradeoff between fidelity
with the optimal allocation and preemption overhead. In practice,
hiding this preemption overhead is achievable with careful engineering.

We compare this to an ideal baseline that allocates resources to jobs \emph{exactly}
according to their computed allocation. As shown in Figure~\ref{fig:mechanism_vs_ideal},
\system{}'s scheduling mechanism with a round duration of 6 minutes behaves
almost identically to this ideal baseline with a single-GPU trace (behavior with a multi-GPU trace is similar).

\begin{figure}[t]
    \centering
    \includegraphics[width=0.8\columnwidth]{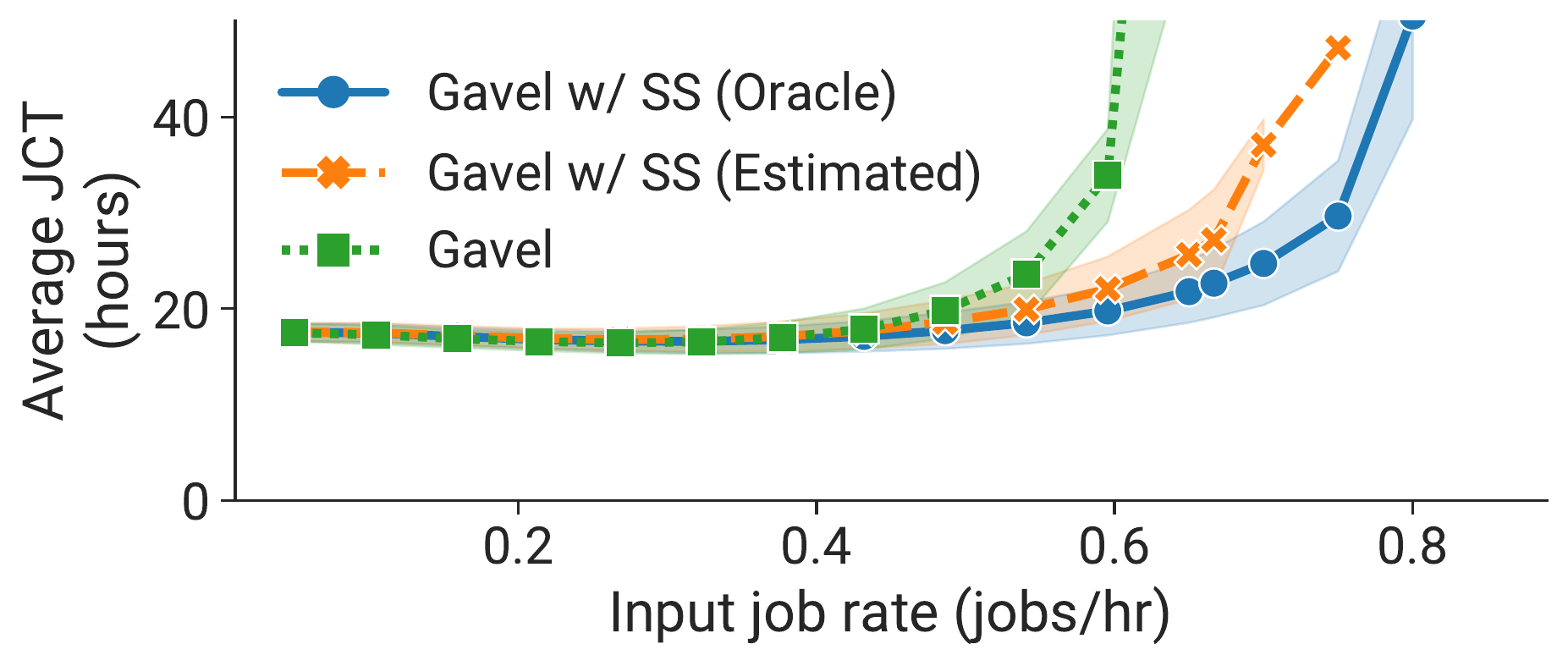}
    \caption{
        \label{fig:throughput-estimation}
        Comparison of SS-aware LAS policy with estimated throughputs, compared
        to the SS-aware with oracle throughputs and LAS without space sharing
        on a heterogeneous 12-GPU cluster.
    }
    \vspace{-0.1in}
\end{figure}

\subsection{Impact of Throughput Estimation} \label{sec:eval_throughput_estimation}

Figure~\ref{fig:throughput-estimation} shows the effect of \system{}'s
throughput estimator on average JCT when using the space sharing-aware LAS policy
compared to the LAS policy without space sharing, and the LAS policy with
space sharing and oracle throughputs. The throughput estimator is able to determine
missing throughputs in an online fashion accurately enough to observe a very
small decrease in average JCT at high load (orange and blue lines).

\section{Related Work} \label{section:related_work}

\paragraph{Existing DNN Training Schedulers.}
Several recent papers have proposed schedulers targeting DNN training workloads.

Gandiva~\cite{xiao2018gandiva} uses time
and space sharing to reduce queuing delay and improve resource utilization,
but does not specify an explicit scheduling policy.
It uses a profiling-based methodology to determine whether to
co-locate jobs on an accelerator. However, it does not incorporate
model performance data (isolated or co-located performance) into
its scheduling policy, resorting to random exploration of job combinations
until a combination that improves performance is found,
and does not support configurable objectives.

Tiresias~\cite{gu2019tiresias} and Themis~\cite{mahajan2020themis} use different
objectives to achieve multi-job fairness. However, both do not incorporate
jobs' affinities for different accelerator types in their scheduling objectives,
and have scheduling mechanisms strongly coupled with the target policy, making
it hard to support other more sophisticated policies like multi-level fairness.

AlloX~\cite{le2020allox} and
$\text{Gandiva}_{\text{fair}}$~\cite{chaudhary2020balancing} are recent DNN
schedulers that do consider worker and model heterogeneity. However, both only
work for single policies (average job completion time for AlloX, max-min
fairness for $\text{Gandiva}_{\text{fair}}$). Moreover,
$\text{Gandiva}_{\text{fair}}$ uses a second-price auction mechanism to improve
the performance of a heterogeneity-agnostic max-min fairness scheme, but does
not provide guarantees as to the optimality of the final allocation. On the
other hand, \system{} formalizes each policy as an optimization problem, and
can provide a guarantee that the returned solution is ``optimal''. \system{}
is also able to support more sophisticated policies such as multi-level fairness.

\paragraph{Traditional Cluster Schedulers.} Traditional schedulers such as
Mesos~\cite{hindman2011mesos}, Borg~\cite{verma2015large},
TetriSched~\cite{tumanov2016tetrisched}, and YARN~\cite{vavilapalli2013apache}
support workloads with fixed heterogeneous resource requests, but do not
reason about the diverse performance characteristics of jobs across accelerators.
Mesos and YARN do not reason about interchangeable resource types that
can run the same computation: for example, Mesos's DRF multi-resource sharing
policy~\cite{ghodsi2011dominant} decides how to give jobs allocations of
distinct resource types, such as RAM and CPUs, but assumes that each job has
declared which resources it needs to use and in what ratio (unlike our case,
where accelerators display heterogeneous performance behavior).

The multi-interchangeable resource allocation (MIRA) problem~\cite{sun2019fair}
also introduces the notion of effective throughput similar to \system{}, but does
not demonstrate how this can be used to specify policies as optimization
problems, does not consider performance optimizations like
space sharing and placement sensitivity, and does not discuss how computed
allocations can be realized on physical resources.

Omega~\cite{schwarzkopf2013omega}, Apollo~\cite{boutin2014apollo}, and
Hydra~\cite{curino2019hydra} are schedulers that
take into account the fact that the target workload shows heterogeneity in
the number and duration of constituent tasks. However, tasks
largely take the same time on different CPUs, and heterogeneity in memory
capacities only impacts the number and size of tasks that can be placed on a
server. In our work, the compute devices themselves are interchangeable with
sometimes large performance differences, and policies decide the time fractions
of resources each job should receive while optimizing for various end objectives.

\paragraph{Dynamic Performance Estimation.}
As detailed in \S\ref{sec:implementation}, Gavel uses the approach proposed
by Quasar~\cite{delimitrou2014quasar} to estimate co-located job performance
online. In particular, Gavel uses a mix of profiling and matrix completion to
compute a ``fingerprint'' against a set of reference models profiled offline.
In this work, we show that the techniques used by Quasar can be successfully
applied to this new setting.

\section{Conclusion} \label{section:conclusion}

In this paper, we proposed \system{}, a heterogeneity-aware cluster scheduler
that is able to optimize for many high-level metrics like fairness, makespan,
and cost. \system{} demonstrates how existing policies can be expressed as
optimization problems, and extends these policies to be heterogeneity-aware.
\system{} then uses a decoupled round-based scheduling mechanism to ensure that
the computed optimal allocation is realized.
\system{}'s heterogeneity-aware policies improve end objectives both on a physical
and simulated cluster. It can support a higher maximum cluster load, while improving
objectives such as average
job completion time by 3.5$\times$, makespan by 2.5$\times$, and cost by 1.4$\times$.

\bibliography{gpusched}
\bibliographystyle{abbrv}

\newpage
\appendix
\section{Appendix}

In this section, we describe how more complicated optimization problem formulations
can be solved in practice, and present additional results.

\subsection{Policies requiring multi-step solutions}\label{app:sec-policy_extensions_multistep}

\paragraph{Minimize makespan.}
The makespan policy tries to allocate resources to jobs in such a way that all active jobs
are completed as soon as possible. To compute an allocation for this policy,
we can binary search for the smallest makespan $M$ that satisfies the following
constraints,
\begin{eqnarray}
& \text{num\_steps}_m \leq \text{throughput}(m, X) \cdot M & \forall m \nonumber \\
& 0 \leq X_{mj} \leq 1 & \forall (m,j) \nonumber  \\
& \sum_j X_{mj} \leq 1 & \forall m \nonumber \\
& \sum_m X_{mj} \leq 1 & \forall j \nonumber
\end{eqnarray}

\paragraph{Identifying bottleneck jobs in fairness policy.}
Solving a max-min fairness policy, such as LAS or hierarchical fairness, results in an allocation
that satisfies fairness metrics but may underutilize resources in scenarios where the bottlenecked job's
throughput is matched by other jobs without using all available resources.
Identifying bottleneck jobs after an iteration of a fairness policy computation
can be done by solving a mixed-integer linear program. The binary integer
variable $z_m$ is set to 1 when job $m$'s scaled effective throughput can
be improved without causing any other job's scaled effective throughput to drop
below the minimum computed in the previous iteration of the policy's LP. We identify all jobs which are
stuck as $\{m: z_m = 0\}$ by computing an allocation that maximizes the sum of
all $z_m$:
$$\text{Maximize}_X \sum_{\{m: p_m > 0\}} z_m$$
\noindent subject to:
\begin{eqnarray}
& z_m = \begin{cases}
1 & \text{ if } \text{throughput}(m,X) > \text{throughput}(m, X^\text{prev}) \nonumber\\
0 & \text{otherwise} \nonumber
\end{cases}
\end{eqnarray}

The conditional constraint on $z_m$ can be expressed as two linear inequalities:
\begin{flalign}
& \text{throughput}(m,X^\text{prev}) < \text{throughput}(m,X) + Y(1 - z_m) \nonumber\\
& \text{throughput}(m,X^\text{prev}) \geq \text{throughput}(m,X) - Y z_m \nonumber
\end{flalign}
where $Y$ is a sufficiently large number such that it is not an active
constraint, such as the max throughput of the job.

\subsection{Additional Results}\label{app:sec-additional_results}
\begin{figure}[t]
  \centering
  \includegraphics[width=0.6\linewidth]{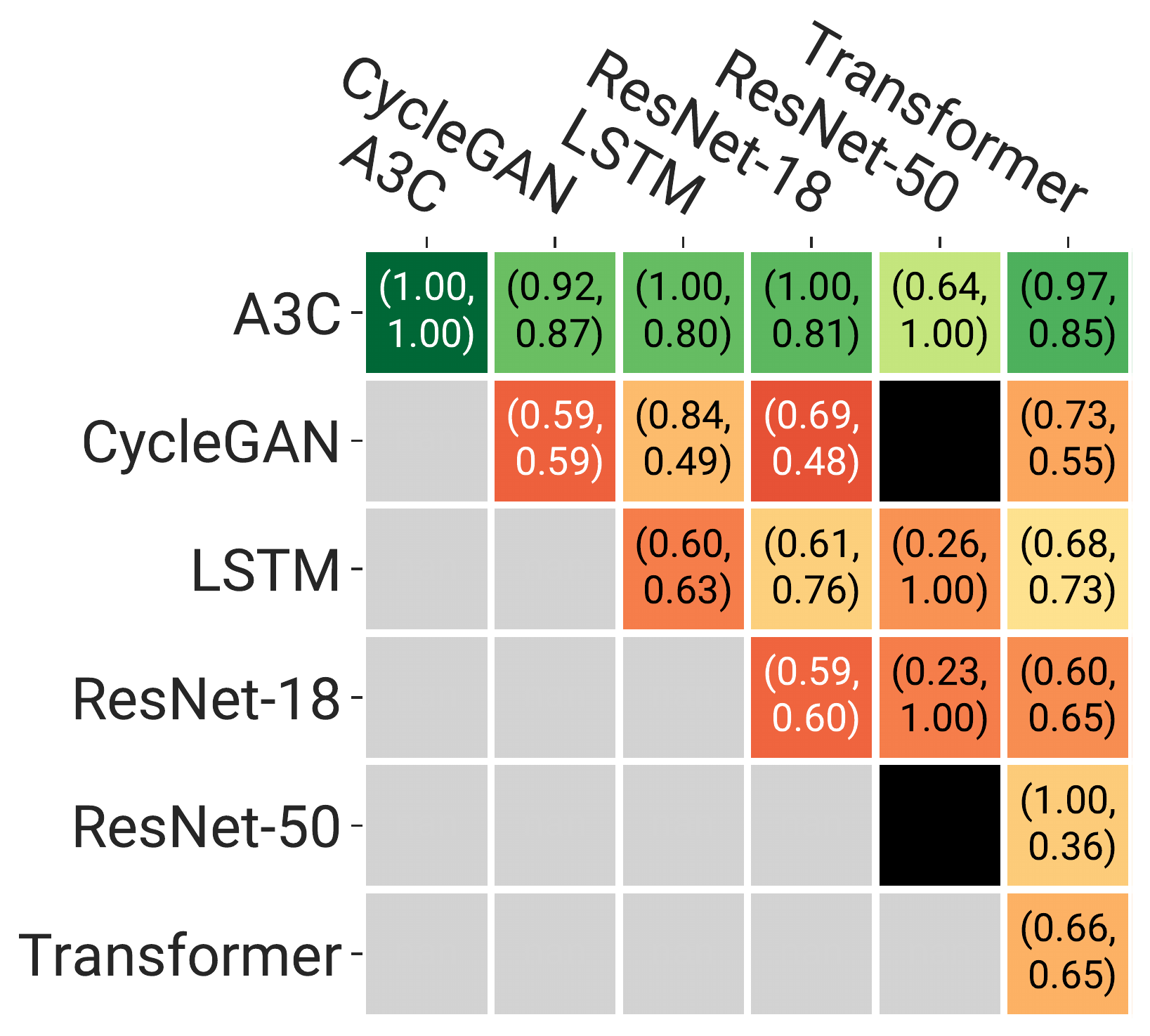}
  \caption{Performance of several DNN models when run concurrently on a single
  P100 GPU. The cell at row $i$ and column $j$ reports the normalized throughput
  (iterations/second) achieved by co-located models $i$ and $j$.
  Throughputs are normalized with respect to the throughput achieved by each model when
  run in isolation. Black squares show jobs that cannot co-locate due to memory constraints.}
  \label{fig:heatmap}
\end{figure}

Figure~\ref{fig:heatmap} shows the performance of colocated DNN
models on a single NVIDIA P100 GPU, presented as a heat map.

\begin{figure}
    \center
    \begin{subfigure}[b]{\columnwidth}
        \includegraphics[width=0.85\columnwidth]{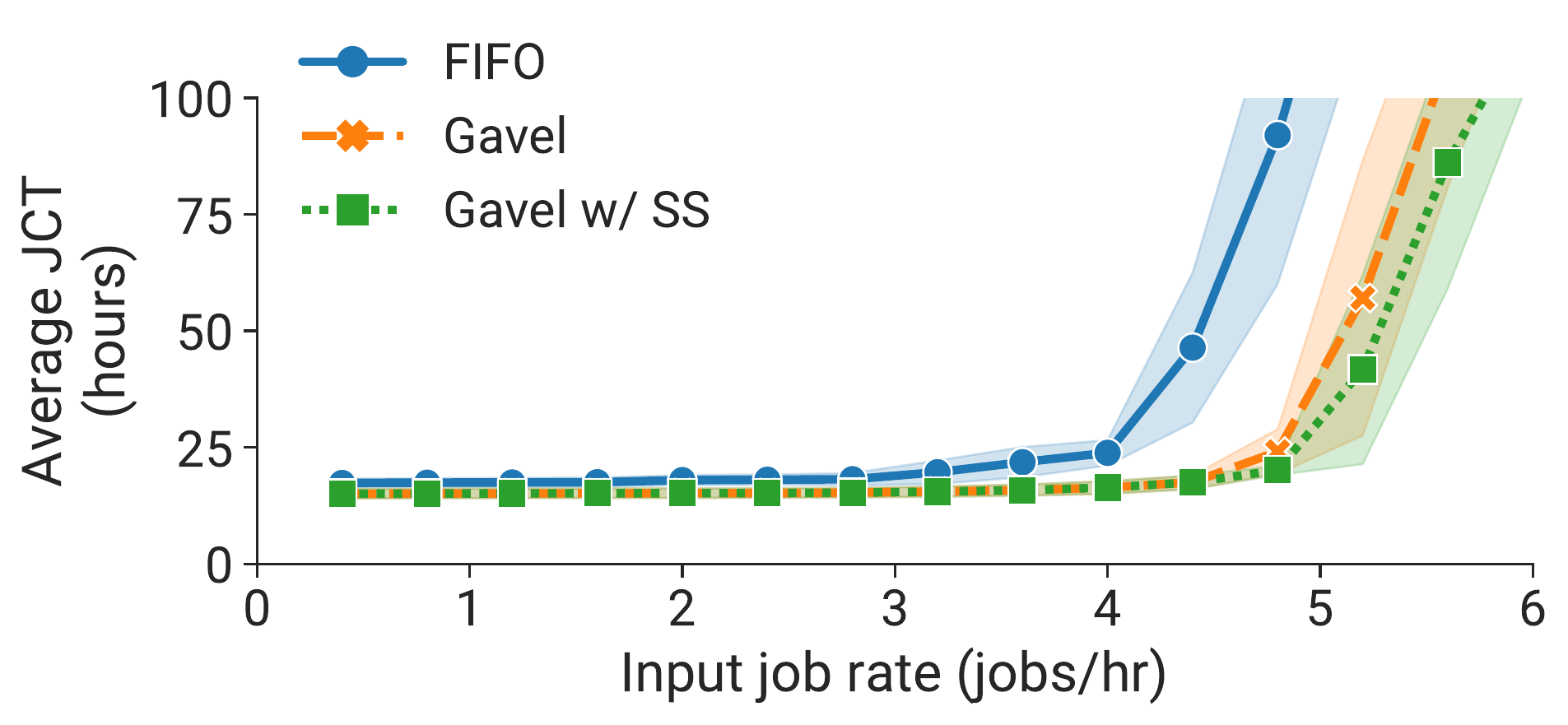}
        \caption{Average job completion time vs. cluster load.}
    \end{subfigure}
    \begin{subfigure}[b]{\columnwidth}
        \includegraphics[width=0.85\columnwidth]{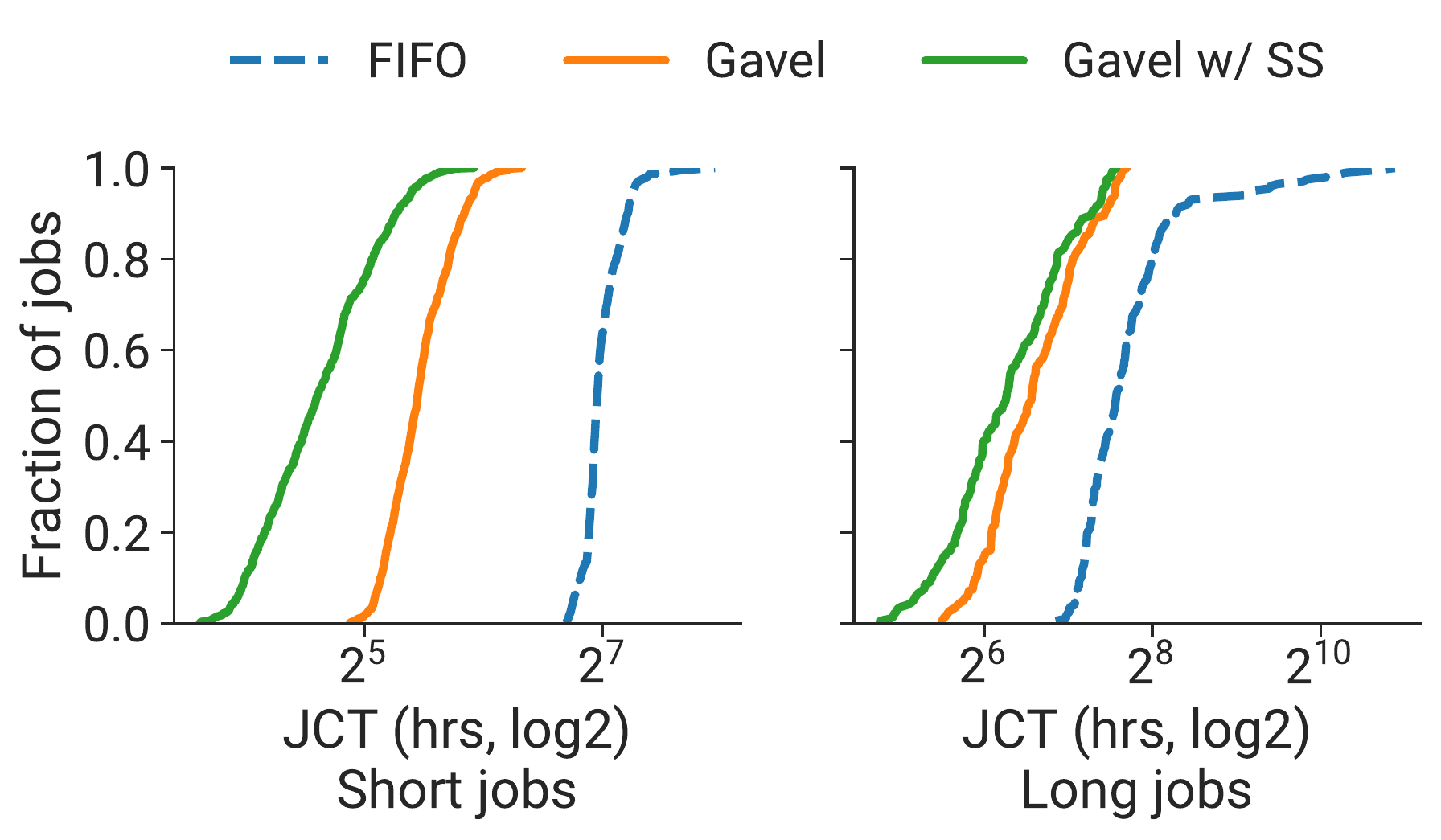}
        \caption{CDF of job completion times.}
    \end{subfigure}
    \caption{
        Comparison of heterogeneity-agnostic FIFO policy
        to a heterogeneity-aware FIFO policy, on the simulated cluster and
        continuous-single trace.
        \label{fig:fifo_single_gpu}
    }
\end{figure}

\begin{figure}
    \center
    \begin{subfigure}[b]{\columnwidth}
        \includegraphics[width=0.85\columnwidth]{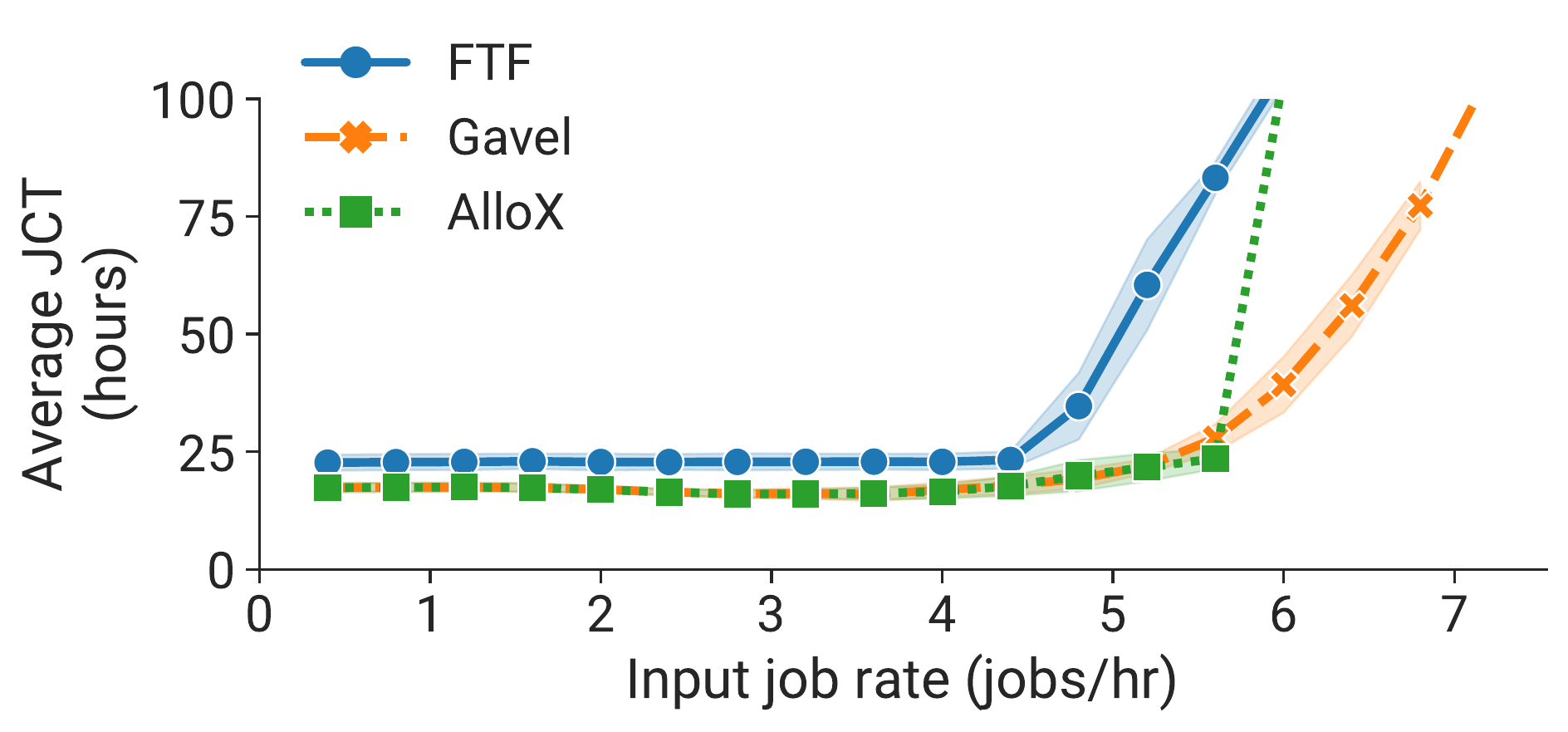}
        \caption{Average job completion time vs. cluster load.}
    \end{subfigure}
    \begin{subfigure}[b]{\columnwidth}
        \includegraphics[width=0.85\columnwidth]{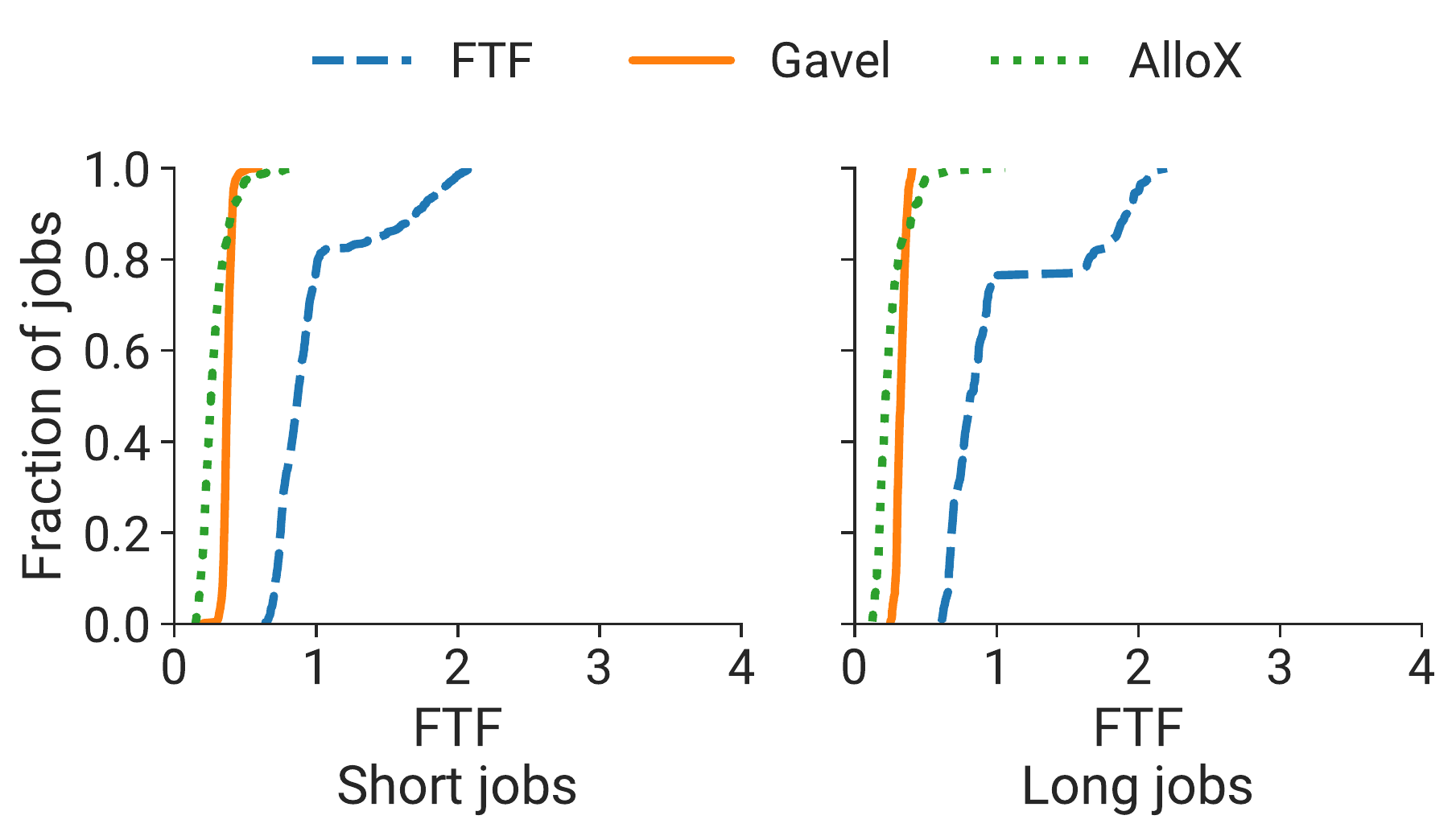}
        \caption{CDF of job completion times.}
    \end{subfigure}
    \caption{
        Comparison of heterogeneity-agnostic finish time fairness (FTF) policy
        to a heterogeneity-aware FTF policy, on the simulated cluster and
        continuous-single trace.
        \label{fig:ftf_single_gpu}
    }
\end{figure}

\begin{figure}
    \center
    \begin{subfigure}[b]{\columnwidth}
        \includegraphics[width=0.85\columnwidth]{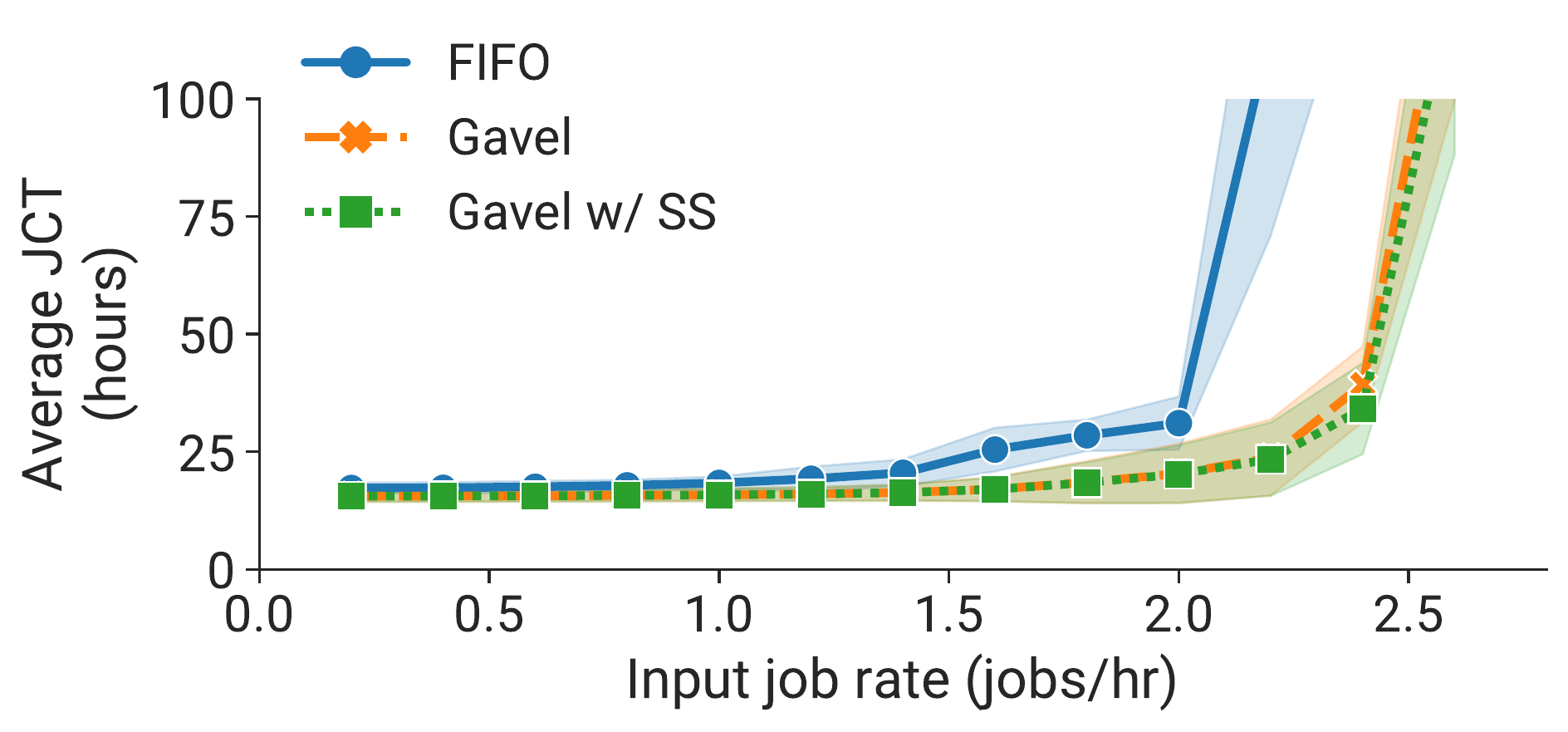}
        \caption{Average job completion time vs. cluster load.}
    \end{subfigure}
    \begin{subfigure}[b]{\columnwidth}
        \includegraphics[width=0.85\columnwidth]{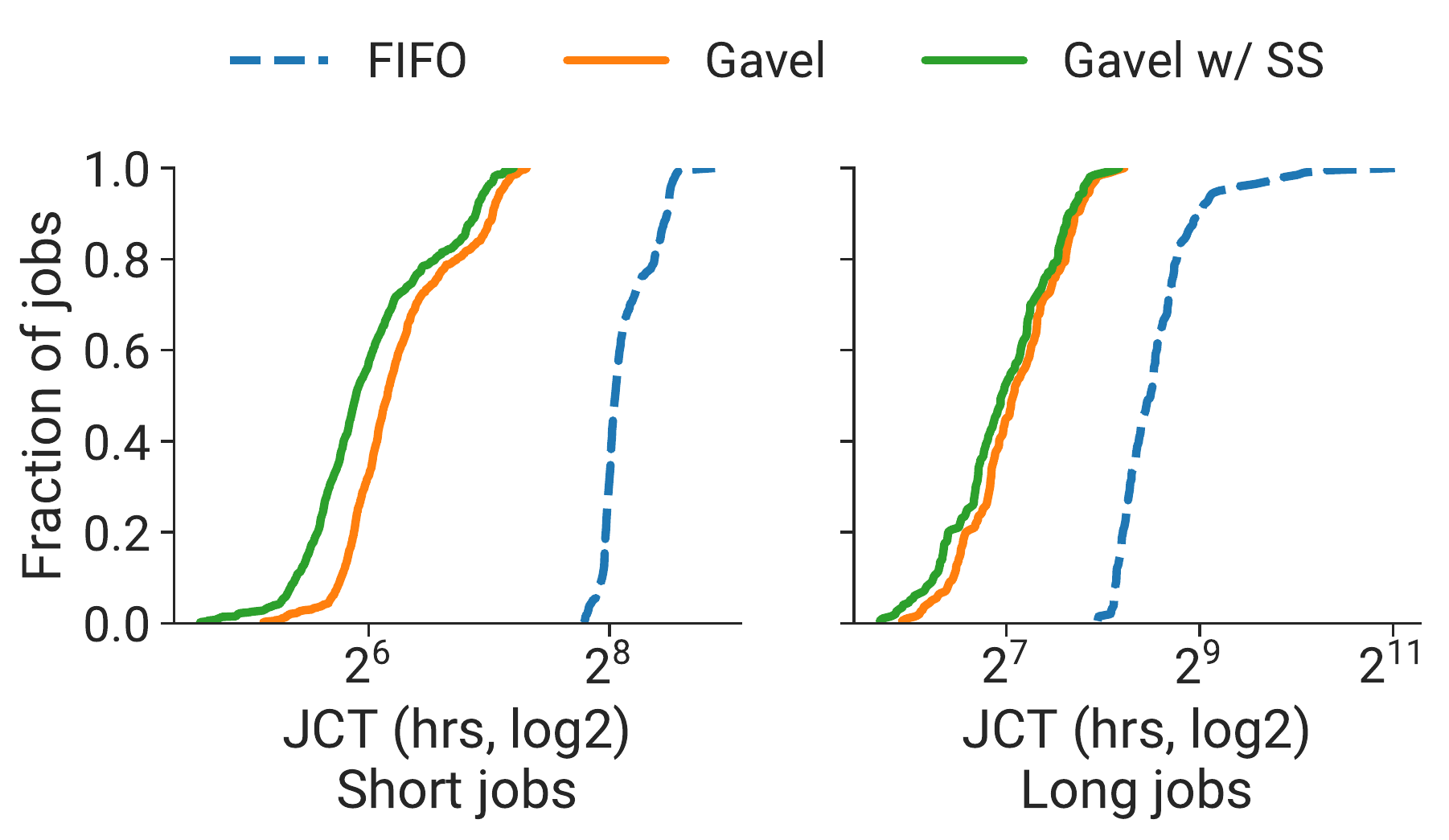}
        \caption{CDF of job completion times.}
    \end{subfigure}
    \caption{
        Comparison of heterogeneity-agnostic FIFO policy
        to a heterogeneity-aware FIFO policy, on the simulated cluster and
        continuous-multiple trace.
        \label{fig:fifo_multi_gpu}
    }
\end{figure}

\begin{figure}
    \center
    \includegraphics[width=0.85\columnwidth]{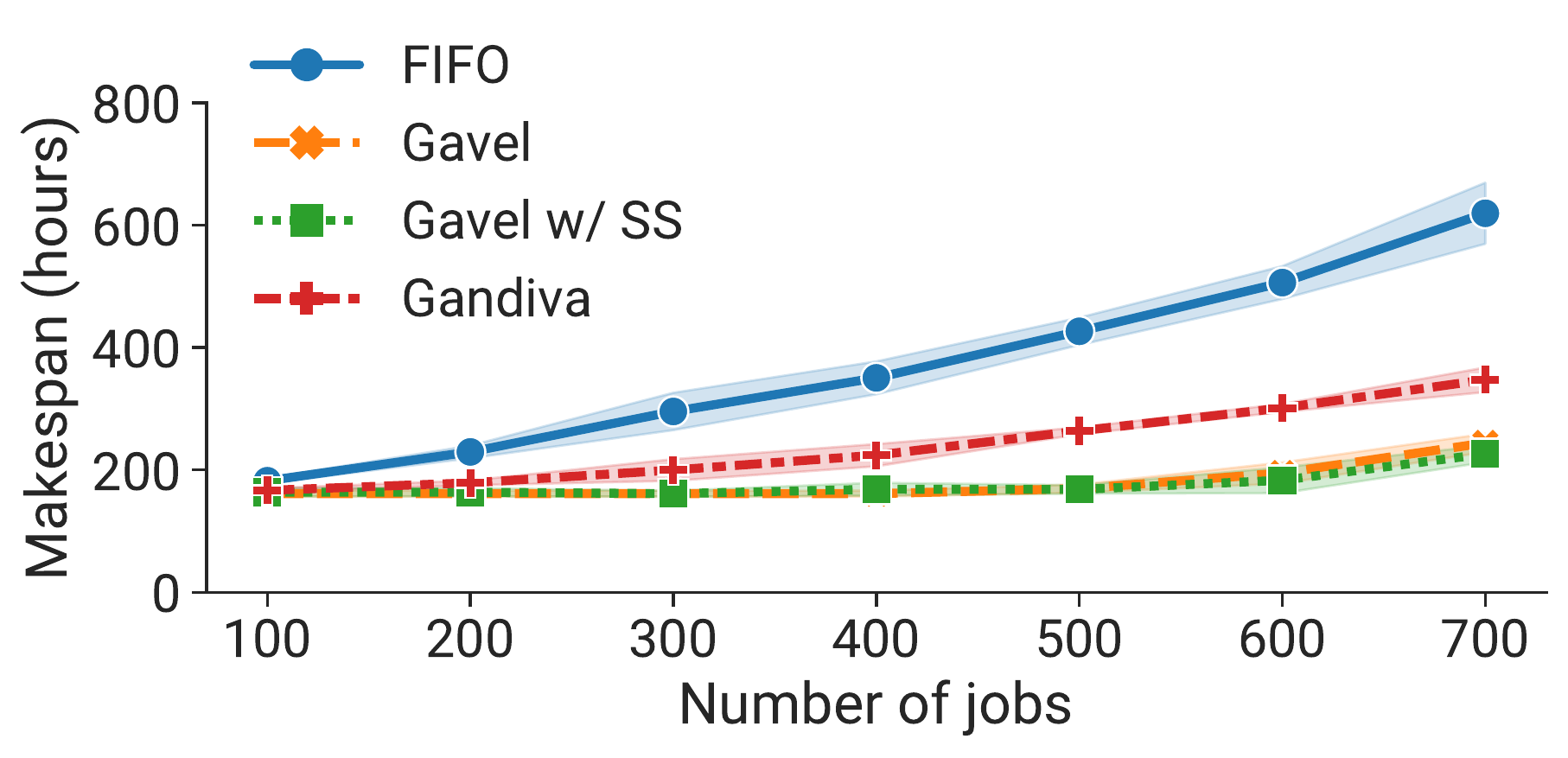}
    \caption{
        Comparison of heterogeneity-agnostic makespan policy
        to a heterogeneity-aware makespan policy, as well as FIFO and Gandiva
        baselines on the simulated cluster and the static-multiple trace.}
    \label{fig:makespan_multi_gpu}
\end{figure}

\begin{figure}
   \center
   \includegraphics[width=0.85\columnwidth]{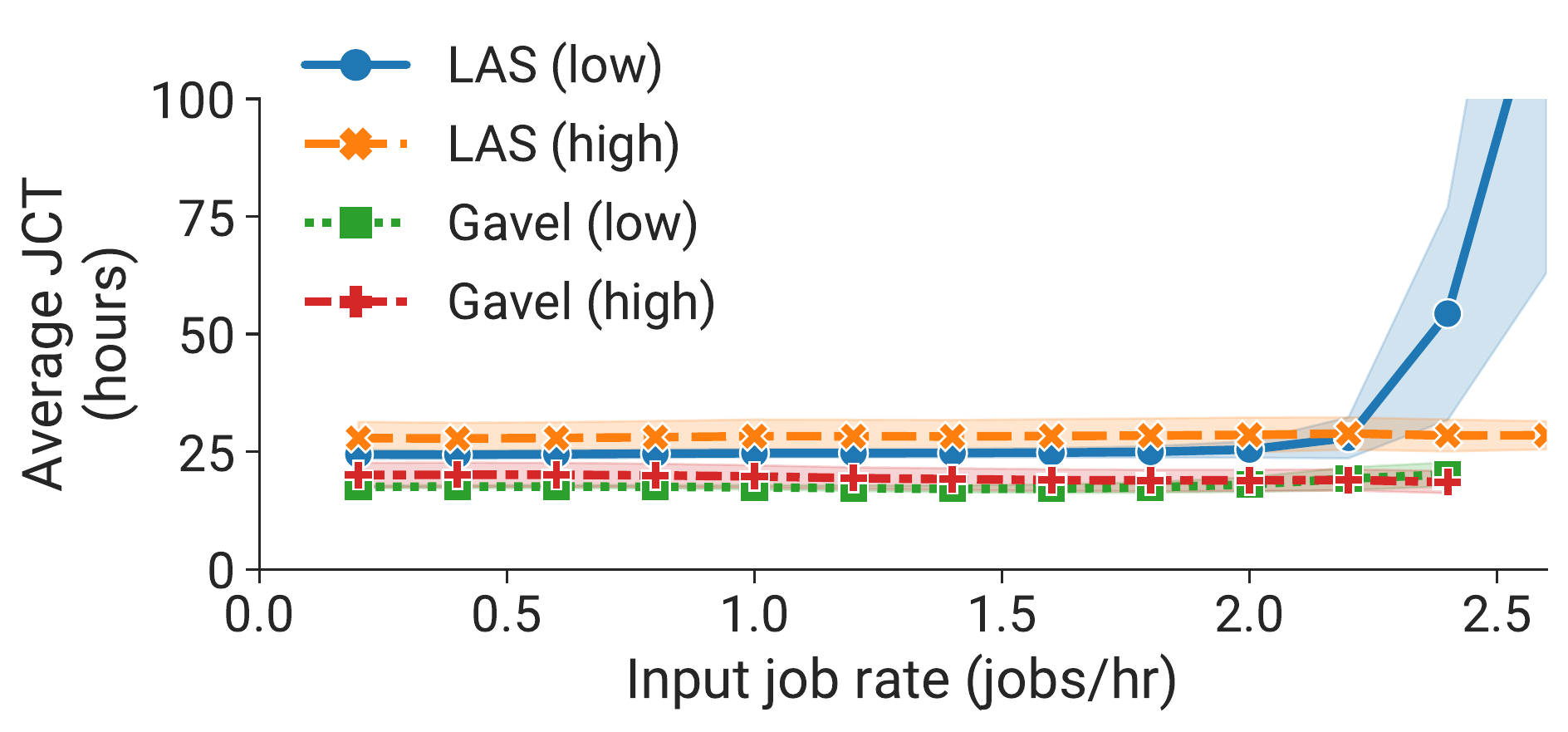}
   \caption{
       Comparison of heterogeneity-agnostic LAS policy
       to a heterogeneity-aware LAS policy,
       on the simulated cluster, continuous-multiple trace, with 20\% of jobs
       with high priority.}
   \label{fig:las_priorities_multi_gpu}
\end{figure}

Figures~\ref{fig:fifo_single_gpu}, \ref{fig:ftf_single_gpu}, \ref{fig:fifo_multi_gpu},
\ref{fig:makespan_multi_gpu}, and \ref{fig:las_priorities_multi_gpu} show a
comparison of vanilla, heterogeneity-agnostic policies with their
heterogeneity-aware counterparts on single-worker and multi-worker traces.
We see across the board that \system{}'s heterogeneity-aware policies allow the
cluster to support higher load (in terms of average job input rate), and also
reduce end objectives.

\begin{figure}
    \centering
    \includegraphics[width=0.85\columnwidth]{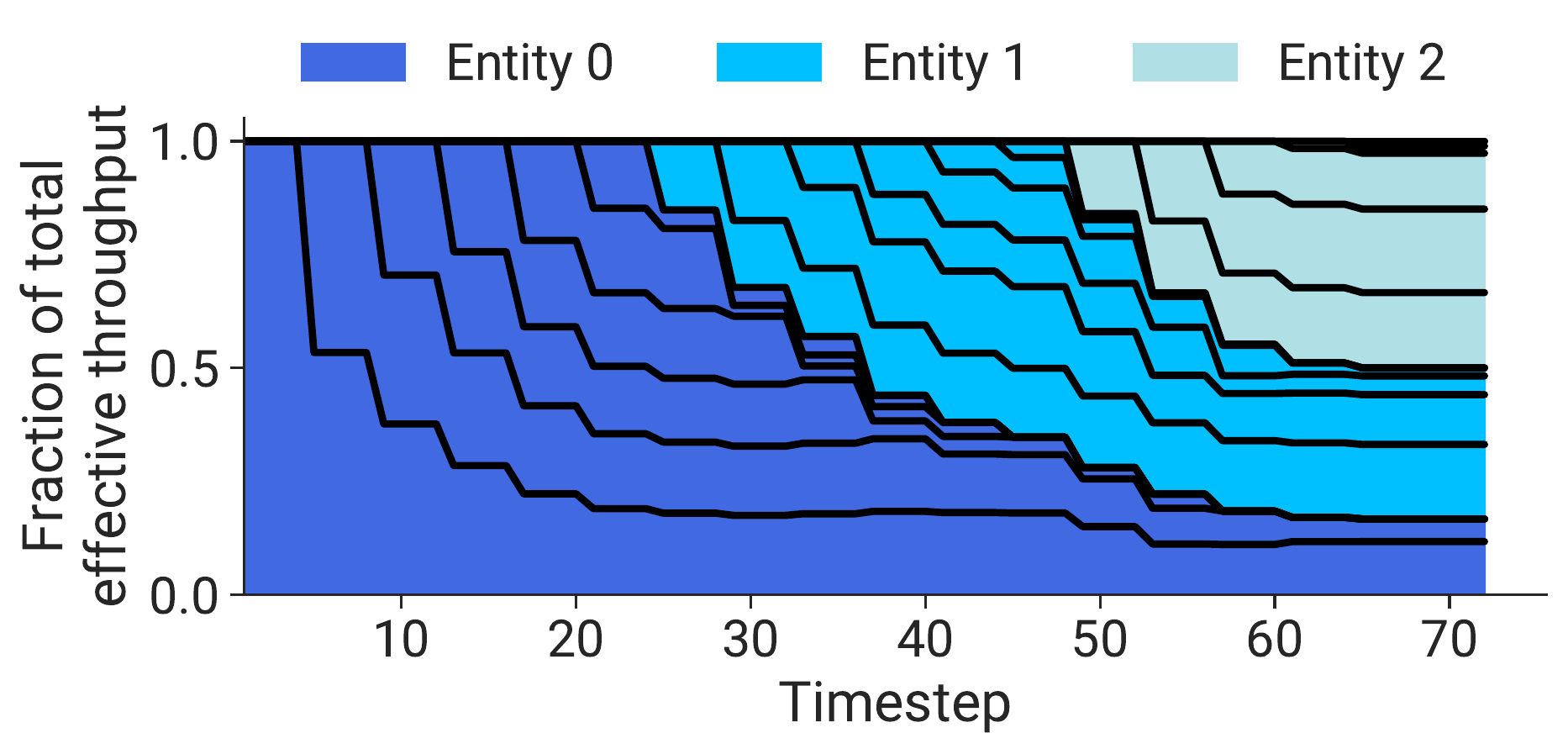}
    \caption{
        \label{fig:hierarchical_fairness_fifo}
        Timeline showing behavior of a hierarchical policy (fairness as top-level
        policy, FIFO as bottom-level policy) with time.
        A job is added every 4 timesteps. The first 6 jobs belong to entity 0,
        the next 6 jobs belong to entity 1, and the last 6 jobs belong to
        entity 2. Entity 0 has a weight of 1, entity 1 has a weight of 2, and
        entity 2 has a weight of 3. Real throughputs are used, and jobs are
        sampled from those shown in Table~\ref{table:model_list}.
    }
\end{figure}

Figure~\ref{fig:hierarchical_fairness_fifo} shows the behavior of a multi-level
fairness policy that combines a higher-level fairness policy with a low-level
FIFO policy. The widths of the colored bands are in the ratios of the weights
of each entity; within an entity, the effective throughputs of jobs are ordered
in a way that respects the arrival order of jobs within that entity. All jobs
within low-weight entities do not receive resources under high load.

\end{document}